\documentclass[%
reprint,
superscriptaddress,
 amsmath,amssymb
 aps,
prb,
floatfix,
]{revtex4-2}
\allowdisplaybreaks
\usepackage{xcolor}
\usepackage{graphicx}
\usepackage{dcolumn}
\usepackage{bm}
\usepackage{hyperref}
\usepackage{circledsteps} 
\usepackage{mathtools} 
\usepackage{amssymb} 
\usepackage{mathrsfs} 
\usepackage{subfigure} 
\newcommand{\CR}{\hat a^\dagger}
\newcommand{\AN}{\hat a}
\newcommand{\CRf}{\hat c^\dagger}
\newcommand{\ANf}{\hat c}
\newcommand{\defeq}{\vcentcolon=}
\newcommand{\eqdef}{=\vcentcolon}
\newcommand{\ham}{\hat{\mathcal{H}}}
\newcommand{\Sum}{\sum\limits}

\newcommand{\eqsec}{\overset{\CircledTop{2}}{=}}
\newcommand{\N}{\mathbb{N}}  
\newcommand{\hamdens}{\hat{\mathscr{H}}}

\begin{document}
\title{Operator Valued Flow Equation Approach to the Bosonic Lattice Polaron: Dispersion Renormalization Beyond the Fröhlich Paradigm}

\author{Jan-Philipp Christ}
\affiliation{%
Faculty of Physics, Arnold Sommerfeld Centre for Theoretical Physics (ASC),
Ludwig-Maximilians-Universität München, Theresienstr. 37, 80333 München, Germany
}%
\author{Pit Bermes}
\affiliation{%
Faculty of Physics, Arnold Sommerfeld Centre for Theoretical Physics (ASC),
Ludwig-Maximilians-Universität München, Theresienstr. 37, 80333 München, Germany
}%
\author{Fabian Grusdt}
\affiliation{%
Faculty of Physics, Arnold Sommerfeld Centre for Theoretical Physics (ASC),
Ludwig-Maximilians-Universität München, Theresienstr. 37, 80333 München, Germany
}%

\date{\today}

\begin{abstract}
    We consider the ground state properties of a lattice Bose polaron, a quasiparticle arising from the interaction between an impurity confined to an optical lattice and a surrounding homogeneous Bose-Einstein condensate hosting phononic modes. We present an extension of Wegner's and Wilson's flow equation approach, the operator valued flow equation approach, which allows us to calculate the renormalized dispersion of the polaron and assess the role of two-phonon scattering processes on the dispersion. The results obtained in this way are compared to a variational mean-field approach. We find that in certain impurity phonon interaction regimes the shape of the dispersion is significantly altered by the inclusion of two-phonon scattering events as opposed to only single-phonon scattering events. Moreover, our results predict that a polaronic bound state may emerge, which is not present in Fröhlich-type models that only consider single-phonon scattering events. 
\end{abstract}

\maketitle

\section{Introduction}
\label{sec:introduction}
Since its proposal by \citet{Landau:1933asb} to describe the interaction of an electron with the atoms in a crystal, the polaron concept has attracted considerable attention among condensed matter physicists for its extensibility to the more general setting where an impurity is dressed by excitations of a surrounding medium. To give just a few examples, these can be, as in the original setting, phonons of crystal vibrations \cite{Fröhlich,2015devreese}, spin excitations \cite{PhysRevB.39.12232,PhysRevB.39.6880,Koepsell2019} or Bogoliubov excitations of a Bose-Einstein condensate (BEC) \cite{PhysRevB.80.184504,Mostaan_Bose_Polaron_Review_2024,PhysRevLett.93.120404} which we consider here. \par
Thus, ultracold quantum gases can provide a platform to study the fundamental polaron concept in a particularly pure and isolated setting. In this context, optical lattices can be used to experimentally implement different potential landscapes \cite{bloch2005ultracold}. Notably, the realization of atom-species-selective optical lattices    has opened the possibility to investigate settings where an impurity atom is confined to a lattice whilst immersed in a spatially homogeneous BEC \cite{PhysRevA.76.011605,LeBlanc_2007}. \par 
On the theoretical side, most descriptions of lattice polarons are based on the Hamiltonian originally introduced by \citet{Fröhlich}. The excitations surrounding the impurity are identified as quasiparticles and the Fröhlich Hamiltonian describes their energies as well as the scattering of single quasiparticles off the impurity. However, it has been shown that for an accurate description of Bose polarons two-phonon scattering terms and, in the strong-coupling regime, even phonon-phonon interactions cannot be neglected \cite{PhysRevA.88.053632,Grusdt_2017,PhysRevLett.117.113002,mostaan2023unifiedtheorystrongcoupling,10.21468/SciPostPhys.16.3.067}.\par
While the ground state and dynamical properties of lattice polarons have been analyzed at the Fröhlich level \cite{Grusdt_2014_Bloch}, open questions on the effect of two-phonon scattering terms for their theoretical description remain. The purpose of this article is to consider a system where an impurity on a quasi one-dimensional optical lattice is immersed in a one-dimensional Bose gas. A similar scenario \textit{without} an optical lattice potential has already been subject to both experimental \cite{Catani_2012} and extensive theoretical investigation, see e.g. \cite{Catani_2012,Grusdt_2017,PhysRevA.95.023619,PhysRevA.92.031601,PhysRevA.96.031601,PhysRevResearch.2.033142,10.21468/SciPostPhys.11.1.008,PANOCHKO2019167933}. Moreover, the spectral and quasiparticle properties of a lattice polaron in a two-dimensional BEC of hard-core bosons \cite{Santiago-Garcia_2024} as well as the dynamics of two interacting polarons in a two-dimensional BEC \cite{10.21468/SciPostPhys.14.6.143} have been subject of previous studies. \par
We will work with Bogoliubov theory and include two-phonon terms while neglecting phonon-phonon interactions. We focus our analysis on the ground state properties of the lattice polaron by calculating its energy and renormalized dispersion. This corresponds to calculating the mass renormalization for non-lattice Bose polarons. We will tackle this problem in two different ways: On one hand we use a variational coherent state ansatz to obtain the mean-field dispersion \cite{Grusdt_2014_Bloch,PhysRevLett.117.113002}; on the other hand we introduce an extension of the flow equation approach \cite{WegnerOG_Article,WilsonOG_Article} to approximately diagonalize the Hamiltonian. The method presented here has the promise of being adaptable to several physical problems, in particular to Hamiltonians arising in bosonic impurity problems \cite{Mostaan_Bose_Polaron_Review_2024}. \par
\begin{figure}
\includegraphics[width=0.48\textwidth]{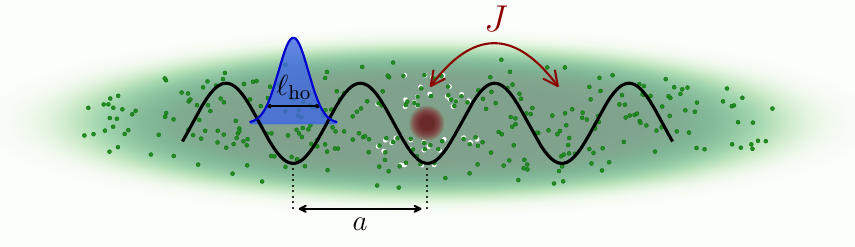}
\caption{A single impurity (red) is immersed in a quasi-one dimensional homogeneous BEC (green) while being constrained to an one-dimensional optical lattice (black sinusoidal line). The distance between adjacent lattice sites is $a$, and the hopping parameter $J$ is associated with the required energy cost for the impurity to tunnel between adjacent sites. The Wannier functions (blue) on each lattice site are approximated by Gaussians of width $\ell_{\mathrm{ho}}$. The distortions of the condensate depicted near the impurity illustrate the formation of a polaron.}
\label{fig:setup_sketch}
\end{figure}
This article is organized as follows:
After introducing the theoretical model in \autoref{sec:model}, we will present our results for the renormalized dispersion of the lattice polaron both with and without two-phonon scattering terms in \autoref{sec:results}. We will also address the dependence of the ground state energy on the impurity-boson interaction strength at fixed polaron momentum. In \autoref{sec:formalism}, we will develop our formalism and introduce the previously mentioned extension of the flow equation approach. We will briefly present the idea of the flow equation approach \autoref{sec:formalism:overview}, then make the transition to allow operators to be subject to the flow (\ref{sec:formalism:transition}) and discuss how solving the flow equations can be approached  (\ref{sec:formalism:ansatz},\ref{sec:formalism:solving}) as well as why the breakdown of the flow equations in a certain regime is to be expected (\ref{sec:formalism:breakdown}). This article is then concluded by a summary of the results obtained for the bosonic lattice polaron, a proposal for an experiment to test the validity of these results, and a discussion on further potential applications of the newly developed operator-valued flow equation approach.

\section{The Model}
\label{sec:model}
We consider a one-dimensional system where a single impurity of mass $m_{\mathrm I}$, described by a field operator $\hat\psi(x)$, is immersed into a sea of bosons of mass $m_{\mathrm B}$, described by the field operator $\hat\phi(x)$, with contact interaction of strength $g_{\mathrm{BB}}$. The impurity is confined to a lattice of length $L$ whereas the bosons move in continuous space, a scenario which can be realized optically \cite{LeBlanc_2007}. The boson-impurity interaction strength is $g_{\mathrm{IB}}$, and in natural units the microscopic Hamiltonian of this system reads 
\begin{align}
\ham &= \int\mathrm dx \Bigg\{ \hat\phi^\dagger(x)\left[-\frac{\nabla^2}{2m_B}+\frac{g_{\mathrm{BB}}}{2}\hat\phi^\dagger(x)\hat\phi(x)\right]\hat\phi(x)\nonumber\\ 
&+\hat\psi^\dagger(x)\left[-\frac{\nabla^2}{2m_{\mathrm I}}+V_{\mathrm I}(x)+g_{\mathrm{IB}}\hat\phi^\dagger(x)\hat\phi(x)\right]\hat\psi(x)\Bigg\}. \label{eq: ham_density}
\end{align}
Here we introduced the optical lattice potential, which we take to be of the form 
\begin{align}
V_{\mathrm I}(x)=V_0\sin^2\left(k_0x\right).
\end{align}
$k_0=2\pi/\lambda$ is the wave vector used to create the lattice potential, which is only seen by the impurity. This setting is similar to the one discussed in \cite{Grusdt_2014_Bloch}, except that we consider a fully one-dimensional system here. Moreover, in our following analysis we will include two-phonon scattering terms.\par
The field operator $\hat\psi(x)$ can be expanded in terms of Wannier functions which will be approximated as Gaussians with standard deviation $\ell_{\mathrm{ho}}$ centered around a given lattice site $j$.
Next, we introduce the spacing between adjacent lattice sites $a$ along with $\omega_0=2\sqrt{V_0k_0^2/2m_{\mathrm I}}$ and $\ell_{\mathrm{ho}}=1/\sqrt{m_{\mathrm I}\omega_0}$ in order to relate the optical potential to the Wannier functions.\par
The free impurity is then described by a tight-binding Hamiltonian 
\begin{equation} 
\ham_{\mathrm I}=-J\Sum_j\left(\hat c_{j+1}^\dagger\hat c_j+\mathrm{h.c.}\right),\label{eq: tight-binding_ham}
\end{equation}
with effective hopping $J$ between adjacent lattice sites $j$. Here $\hat c_j^\dagger$ creates an impurity in the Wannier orbital centered around site $j$.\par
The bosons will be treated as an unperturbed homogenius condensate of density $n_0$, speed of sound $c$ and healing length $\xi$ using Bogoliubov theory. Within Bogoliubov theory, the bosonic field operator can be written as $\hat\phi(x)=\sqrt{n_0}+\hat\Phi(x)$ where \begin{align}\hat\Phi(x)&=\frac{1}{\sqrt{2\pi}}\int \mathrm dk\hat\Phi_ke^{-ikx}\nonumber\\&=\frac{1}{\sqrt{2\pi}}\int \mathrm dk\left(u_k\AN_k+v_k\CR_{-k}\right)e^{-ikx}, \label{eq: bog_approx}\end{align} describes fluctuations around the condensate. $\AN_k$ represents a Bogoliubov phonon of momentum $k$ and energy \begin{equation}\omega_k=ck\sqrt{1+\frac12\xi^2k^2}. \label{eq: bog_disp}\end{equation}
Its contribution to the excitation to the condensate is given by the mode functions 
\begin{align}
u_k&=\ \ \frac{1}{\sqrt{2}}\sqrt{\frac{1+k^2\xi^2}{k\xi\sqrt{2+k^2\xi^2}}+1}\\
v_k&=-\frac{1}{\sqrt{2}}\sqrt{\frac{1+k^2\xi^2}{k\xi \sqrt{2+k^2\xi^2}}-1}.
\end{align}
As it will be useful later, we already introduce the dimensionless constant \begin{equation}g_{\mathrm{eff}}=g_{\mathrm{IB}}\sqrt{\frac{n_0}{\xi c^2}}.\end{equation} Depending on whether the impurity-boson interactions are repulsive or attractive, it can take positive or negative values.\par
Combining eqs. \eqref{eq: ham_density}, \eqref{eq: bog_approx}, \eqref{eq: bog_disp}, as well as the Gaussian Wannier functions, gives rise to the following Hamiltonian:
\begin{align}
\ham &=\int \mathrm dk\omega_k \CR_k\AN_k-J\Sum_j\left(\hat c_{j+1}^\dagger\hat c_j+\mathrm{h.c.}\right)\nonumber\\
&+\Sum_j \CRf_j\ANf_j\Bigg[g_{\mathrm{IB}}n_0+\frac{g_{\mathrm{IB}}}{2\pi}\int\mathrm dk v_k^2\nonumber\\ 
&+\int\mathrm dk  \left(\CR_k+\AN_{-k}\right)U_k e^{iajk}\nonumber\\
&+\int\mathrm dk\int\mathrm dk^\prime \CR_k\AN_{k^\prime} V_{k,k^\prime}e^{-iaj(k^\prime-k)}\nonumber\\
&+\int\mathrm dk\int\mathrm dk^\prime \left(\CR_k\CR_{k^\prime} W_{k,k^\prime}e^{iaj(k+k^\prime)}+\mathrm{h.c.}\right)\Bigg]. \label{eq: ham_before_LLP}
\end{align}
Crucially, this Hamiltonian is at most quadratic in impurity or phonon operators. This is because interactions between bosons that are not part of the condensate have been neglected, which is valid in the dilute limit or, alternatively, when the boson-boson coupling $g_{\mathrm{BB}}$ is small.
In eq. \eqref{eq: ham_before_LLP}, we introduced the coefficients
\begin{align}
U_k&=g_{\mathrm{IB}}\sqrt{\frac{n_0}{2\pi}}\left(\frac{(k\xi)^2}{2+(k\xi)^2}\right)^{1/4}e^{-k^2\ell_{\mathrm{ho}}^2/4}\\
V_{k,k^\prime}&=\frac{g_{\mathrm{IB}}}{2\pi}\left(u_ku_{k^\prime}+v_kv_{k^\prime}\right)e^{-(k^\prime-k)^2\ell_{\mathrm{ho}}^2/4}\\
W_{k,k^\prime}&=\frac{g_{\mathrm{IB}}}{4\pi}\left(u_kv_{k^\prime}+v_ku_{k^\prime}\right)e^{-(k^\prime+k)^2\ell_{\mathrm{ho}}^2/4}.
\end{align}\par
We will now show that the Hamiltonian can be transformed such that it factorizes into a block-diagonal form
\begin{equation}
\ham = \Sum_{q\in\mathrm{BZ}}\CRf_q\ANf_q \ham_q. \label{eq: Fourier_cq}
\end{equation}
This reduces the problem of diagonalizing the full Hamiltonian to diagonalizing each purely bosonic block $\ham_q$.
In order to bring the Hamiltonian \eqref{eq: ham_before_LLP} in the block-diagonal form \eqref{eq: Fourier_cq}, we define  the Lee-Low-Pines unitary operator, originally proposed by \citet{LLP_Paper}, \begin{equation} \label{eq: LLP_U}
\hat U_{\mathrm{LLP}}=\exp\left(i\Sum_j aj\CRf_j\ANf_j\int\mathrm dk k\CR_k\AN_k\right).
\end{equation}
Applying it to the Hamiltonian transforms the latter into the polaron frame: Noting that $\hat X \defeq \Sum_j aj\CRf_j\ANf_j$ is the position operator of the impurity and that $\hat P\defeq \int\mathrm dk k\CR_k\AN_k$ is the  momentum operator of the phonons, or more formally using the Baker-Campbell-Hausdorff formula, it is clear that $\hat U_{\mathrm{LLP}}$ translates all bosons by $\hat X$ and boosts the impurity's momentum by $\hat P$.
Finally, after applying transformation \eqref{eq: LLP_U}, we see that Hamiltonian \eqref{eq: ham_before_LLP} indeed factorizes in the block-diagonal form \eqref{eq: Fourier_cq}. The momentum blocks are given by the extended lattice Bogoliubov-Fröhlich (BF) Hamiltonian:
\begin{widetext}
\begin{align}
\ham_q&=\int\mathrm d{k}\left[\omega_k\CR_{{k}}\AN_{{k}}+\left(\CR_{{k}}+\AN_{{k}}\right)U_k\right]+\int\mathrm d{k} d{k^\prime}\left[\left(  \CR_{k}\CR_{{k}^\prime}W_{{k},{k}^\prime}+ \mathrm{h.c.}\right)+ \CR_{k}\AN_{{k}^\prime}V_{{k},{k}^\prime}\right]\nonumber\\
&-2J\cos\Big(aq-a\hat P\Big)+g_{\mathrm{IB}}n_0+\frac{g_{\mathrm{IB}}}{2\pi}\int\mathrm dk v_k^2 . \label{eq: ham_q_with_2phon}
\end{align}
\end{widetext}

\section{Results}
\label{sec:results}
In this section, we present our findings for the ground state energy of the bosonic lattice polaron obtained with the flow equation approach. In anticipation of a portion of the forthcoming discussion in \autoref{sec:formalism}, we will open the present section by conceputally introducing the operator valued flow equation approach. We will also briefly discuss the lattice-Lee-Low-Pines mean-field (MF) treatment via a coherent state ansatz, previously applied in Ref. \cite{Grusdt_2014_Bloch}, against which this new method will be compared. \par
We will first investigate the pure Fröhlich-type Hamiltonian in section \ref{sec:Results:Fröhlich}, i.e. the BF Hamiltonian \eqref{eq: ham_q_with_2phon} with $V_{k,k^\prime}=W_{k,k^\prime}=0\ \forall k,k^\prime$. In section \ref{sec:Results:BogFröhlich}, we will take two-phonon scattering terms into account, which will make our model exact at the level of Bogoliubov theory. This corresponds to the full extendend Bogoliubov Fröhlich Hamiltonian \eqref{eq: ham_q_with_2phon}.

\subsection*{Brief Overview of Theoretical Methods}
The flow equation approach \cite{WegnerOG_Article,WilsonOG_Article,kehrein2006flow} can be used to obtain, at least approximately, the spectrum of a given Hamiltonian by applying a continuous series of unitary transformations. The unitary transformations are ordered by the so-called flow parameter $\lambda$. It is common to design the unitary transformation in such a way that the degree to which off-diagonal matrix elements of the Hamiltonian are suppressed depends on how large the difference in the energies of the associated states is.
In essence, this implies that for a given state with a high energy, a greater number of associated off-diagonal matrix elements are rapidly suppressed in comparison to a state with a lower energy.  This is qualitatively similar to the Wilson-Fisher momentum-shell renormalization group (RG) where fast field components, which for typical dispersions also correspond to higher energies, are integrated out first. However, in contrast to Wilsonian RG, the flow equation approach does not integrate out any degrees of freedom and gives access to the full energy spectrum. Moreover, as in RG, terms can be generated in the flow that are not present in the original Hamiltonian (or action). \par
In our case, the BF Hamiltonian \eqref{eq: ham_q_with_2phon} is not purely quadratic in the bosonic operators $\AN_k$, unless the hopping $J\rightarrow 0$ is suppressed. In the latter case, the flow equation approach can be applied straightforwardly and becomes exact.
For non-zero hoppings, $J\neq 0$, the off-diagonal part of the Hamiltonian will become dependent on the bosonic momentum operator $\hat P$ during the flow. The central step in our formalism is therefore to promote all coefficients $\omega,U,W,V$ in the BF Hamiltonian \eqref{eq: ham_q_with_2phon} to flowing operators commuting with the bosonic number operators $\hat n_k$. \par
We will have to make two approximations:
First, terms that are generated in the flow and do not formally correspond to this form will be neglected. Second, only an approximate representation of the flow operators with a finite number of parameters is used. We will use use two different ansätze: \par
The first one will be to write the flowing operators as a truncated Fourier series in $aq-a\hat P$, which can be motivated easily because any continuous function over the Brillouin zone (BZ) can be written as a Fourier series. An operator $\hat O$ will be represented by the coefficients $\{\alpha_n^{(\hat O)}\}_{0\leq n\leq N}$ ($\{\beta_n^{(\hat O)}\}_{0< n\leq N}$) for cosine (sine) terms with argument $n(aq-a\hat P)$; $N\in\N$ is the Fourier series cutoff. \par
The second one will be a heuristic ansatz involving a finite number of particularly chosen basis functions \eqref{eq:heur_ansatz_basis}, which is less generic than the Fourier series treatment and cannot be motivated without additional assumptions about the shape of the dispersion. Specifically, we have designed them so that their Fourier spectrum decreases exponentially with frequency.\par
Our results obtained via the flow equation aproach will be compared with those  from mean field theory. The variational states corresponding to the MF ansatz read \begin{align}
|\psi_q^{\mathrm{MF}}\rangle=\prod_k \exp\left(\alpha_k\CR_k-\alpha_k^* \AN_k\right)|0\rangle,
\end{align}
with variational parameters $\{\alpha_k\}_k$. It is clear that the flow cannot be accurate for $E_{\mathrm{flow}}>E_{\mathrm{MF}}$ because mean field gives a variational upper bound for the energy. This statement can even be strengthened: With the flow equations we will diagonalize the entire family of operators $\ham_q$ simultaneously. However, the approximations made are not dependent on $q$. So if the flow dispersion is higher in energy than the MF dispersion, which sets an upper bound for each $q$ separately, even for just one $q\in\mathrm{BZ}$, this indicates poor performance of the flow equation approach.

\subsection{Fröhlich Hamiltonian} \label{sec:Results:Fröhlich}
Since it will soon become apparent that in the supersonic regime, where the group velocity $v_g = \langle \partial_q \ham(q)\rangle$ exceeds the speed of sound $c$ of the condensate, the flow equations become unreliable even for relatively small interaction strengths, we will present selected dispersions for the supersonic regime and then primarily focus our analysis to the subsonic regime of the polaron. The supersonic regime with $v_g>c$ involves dynamics that are far from equilibrium including Cherenkov-type radiation and strong decoherence effects \cite{Mostaan_Bose_Polaron_Review_2024}, which we will not address in this article.\par
\begin{figure}
\includegraphics[width=0.48\textwidth]{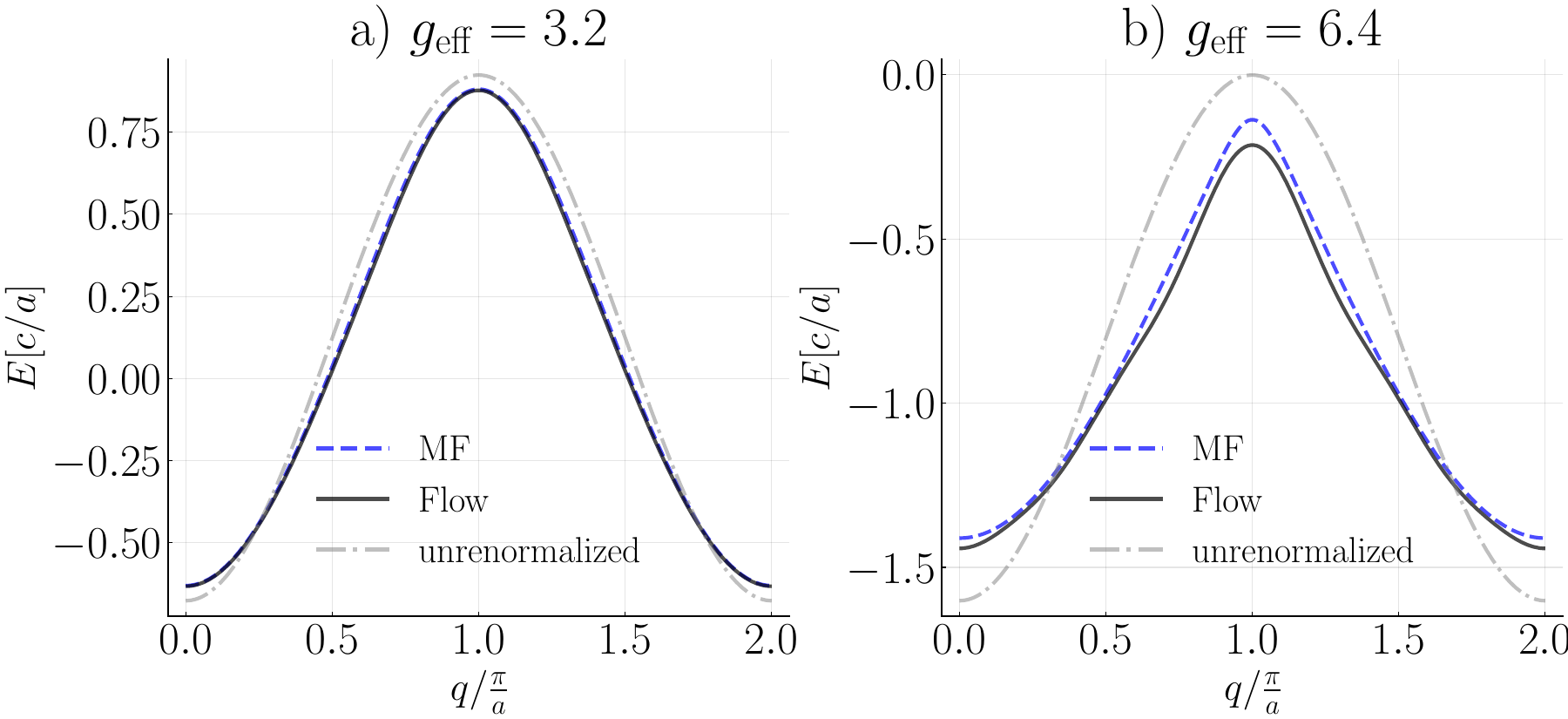}
\caption{Polaron dispersion obtained via MF (blue dashed line) and the flow equations (black solid line) with Fourier cutoff $N=6$ for two different interaction strengths $g_{\mathrm{eff}}$ without the inclusion of two-phonon scattering terms. In both cases, the renormalization of the hopping is most noticeable by the narrowing of the dispersion at the edge of the Brilloin zone ($q=\pm \pi/a$). We also show the unrenormalized dispersion $-2J\cos(aq)$ (grey dash-dotted line), which has been shifted down to the energies of the renormalized dispersion for the purpose of graphical clarity. The hopping was set to $J=0.4 c/a$, the lattice spacing to $a=\xi/5$ and the oscillator length to $\ell_{\mathrm{ho}}=a/\sqrt{2}$. The calculations were performed on a one-dimensional momentum grid of size $N_{\mathrm{grid}}=12$.}
\label{fig:disp_fröhlich_J=0.4}
\end{figure}
\autoref{fig:disp_fröhlich_J=0.4} compares the MF ansatz dispersion to the dispersion obtained via the flow equations (FE). For relatively small interaction $g_{\mathrm{eff}}=3.2$ in \hyperref[fig:disp_fröhlich_J=0.4]{Figure 1a}, MF and the flow equations yield essentially identical results. For larger interation shown in \hyperref[fig:disp_fröhlich_J=0.4]{Figure 1b}, the MF dispersion gets narrower around $q=\pm\pi/a$ at the edge of the BZ. The FE predict a very slight correction to the MF energies while validating the narrowing of the dispersion at the edge of the BZ. The total renormalization of the hopping parameter amounts to around $10\%$ in \hyperref[fig:disp_fröhlich_J=0.4]{Figure 1b}, being slightly stronger for the FE compared to MF. Also, the FE dispersion is strictly lower in energy than the MF dispersion, strongly suggesting that the FE are well-behaved and accurate in this regime.\par

\begin{figure}
\includegraphics[width=0.48\textwidth]{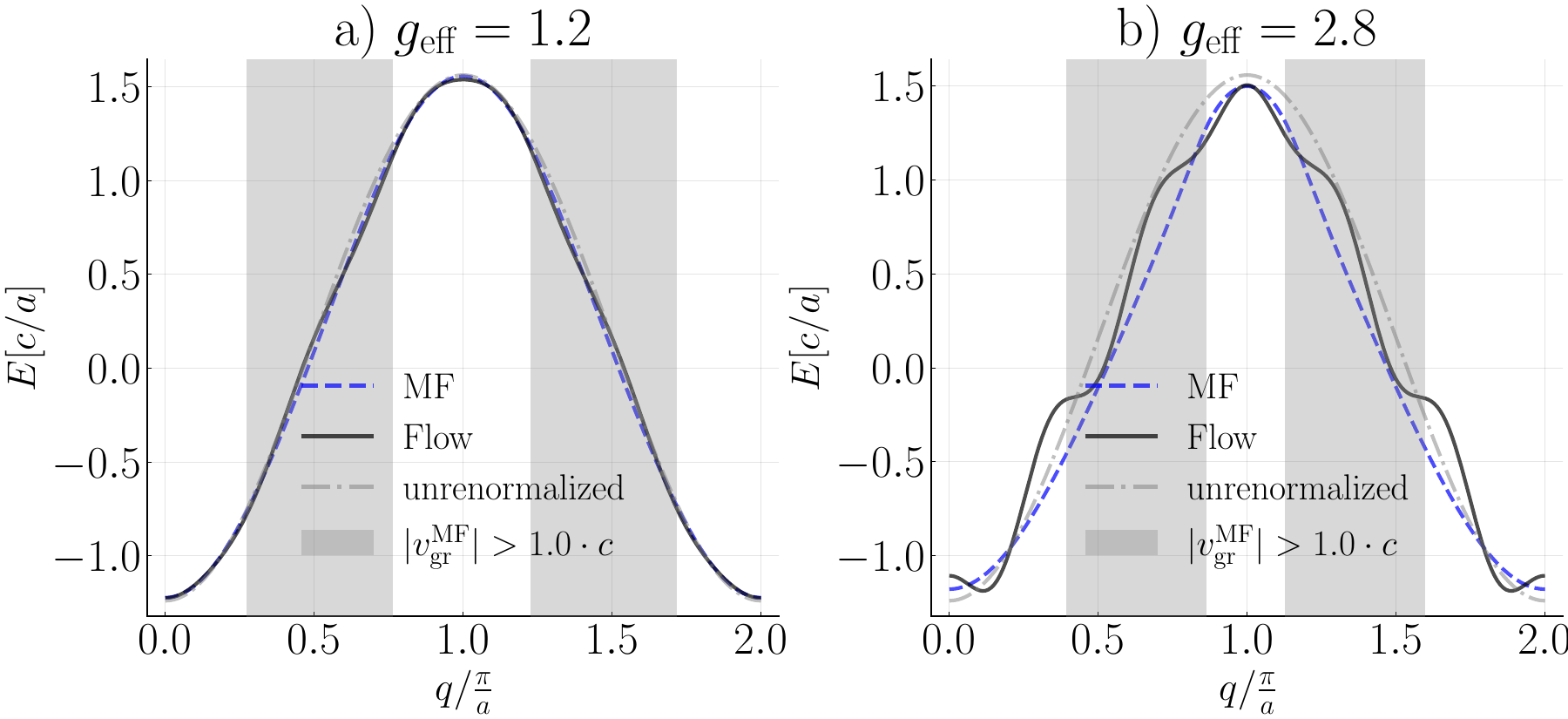}
\caption{Polaron dispersion obtained via MF and the flow equations (Fourier cutoff $N=6$) for two different interaction strengths $g_{\mathrm{eff}}$ in the supersonic regime without the inclusion of two-phonon scattering terms. The hopping was set to $J=0.7 c/a$, which is sufficiently large that there exist $q\in$BZ such that the group velocity exceeds the speed of sound $c$ of the condensate. In the grey shaded area we numerically find $v_{\mathrm{gr}}^{\mathrm{MF}}\geq c$. All system parameters apart from the hopping were chosen exactly as in \autoref{fig:disp_fröhlich_J=0.4}.}
\label{fig:disp_fröhlich_J=0.7}
\end{figure}

In the case of larger hoppings, allowing supersonic bare impurities, the situation is different. \hyperref[fig:disp_fröhlich_J=0.7]{Figure 2a} confirms that for a small range of weak interaction strengths, good agreement between the FE and MF can be obtained even in the supersonic regime. In the case under consideration, the hopping is sufficiently large that the polaron becomes supersonic in the gray-shaded regions centered around $q\sim(\pi\pm\pi/2)/a$. \\
For the slightly larger interaction strength $g_{\mathrm{eff}}=2.8$ considered in \hyperref[fig:disp_fröhlich_J=0.7]{Figure 2b}, the MF dispersion is most substantially renormalized for $q$ within the supersonic (gray-shaded) region, whereas the FE dispersion continues to exhibit a qualitative alignment with the MF dispersion. However, the FE also introduce an oscillatory behaviour on top where high frequency terms in the Fourier ansatz become more dominant than lower frequency terms. Since the FE are energetically above MF for the majority of the BZ, we can state, as discussed above, that the limits of the approximations made have been reached here and the FE are no longer reliable.
Phenomenologically, we observe that MF and FE quantitatively agree if $\alpha_{n+1}^{(H_0)}<\alpha_{n}^{(H_0)}\ \forall 0< n<N$, using notation introduced in \autoref{sec:formalism:ansatz} below. \par
It is based on this observation that we introduce a heuristic ansatz that allows us to no longer represent the dispersion by a truncated Fourier series, but by the appropriately chosen basis functions \eqref{eq:heur_ansatz_basis}. 
\begin{figure}
\includegraphics[width=0.48\textwidth]{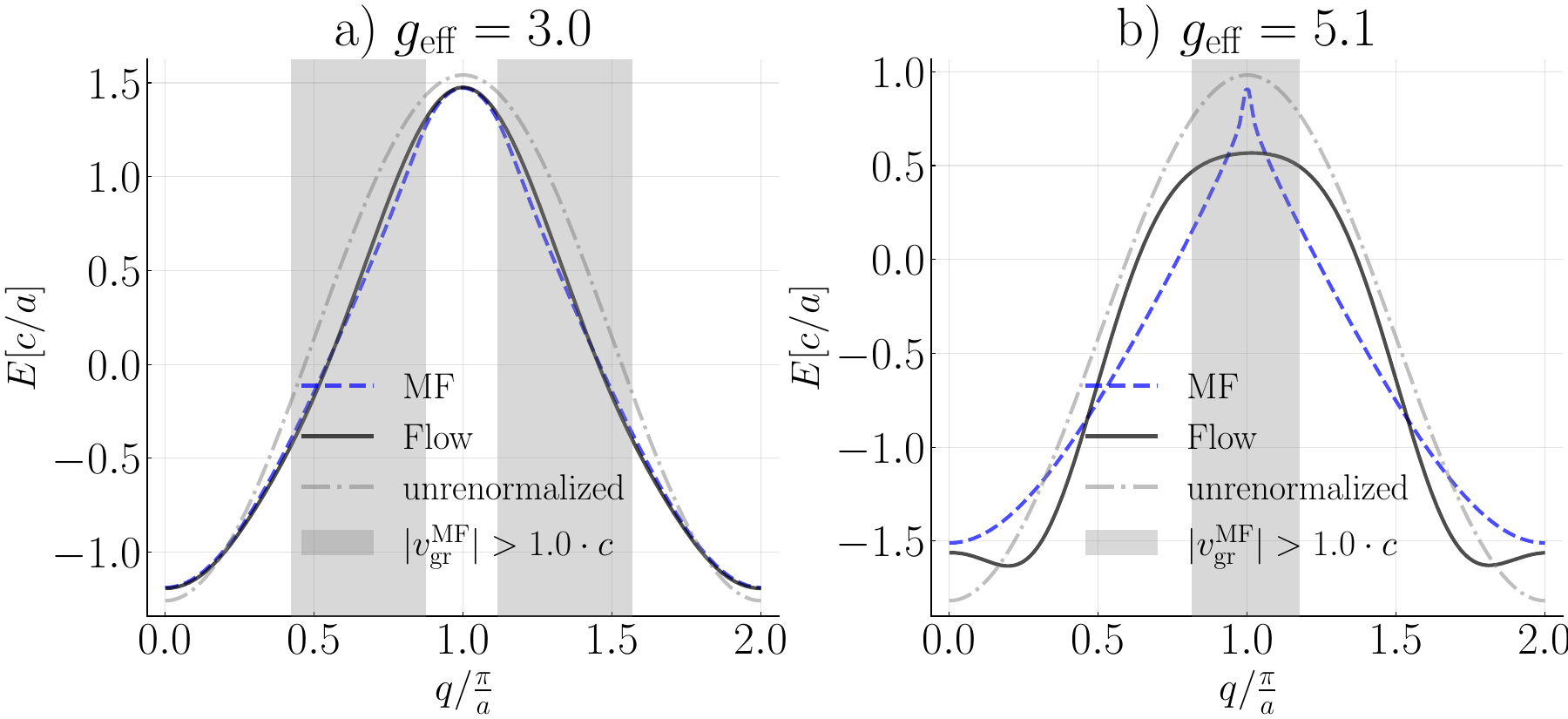}
\caption{Polaron dispersion obtained via MF and the flow equations (heuristic ansatz \eqref{eq:heur_ansatz_basis} with $i\in\{1,2,3,4\}$)  for two different interaction strengths $g_{\mathrm{eff}}$ in the supersonic regime without the inclusion of two-phonon scattering terms. The hopping was set to $J=0.7 c/a$ and the depiction follows that in \autoref{fig:disp_fröhlich_J=0.7}.}
\label{fig:disp_fröhlich_J=0.7_heur}
\end{figure}
Within this ansatz, we obtain a good agreement between FE and MF for a wider range of interactions strenghts, as becomes apparent by comparing \hyperref[fig:disp_fröhlich_J=0.7]{Figure 2b} and \hyperref[fig:disp_fröhlich_J=0.7_heur]{Figure 3a}. Since MF as a variational approach provides an upper bound for the ground state energy for each $q$, \hyperref[fig:disp_fröhlich_J=0.7_heur]{Figure 3b}, where the FE for some phonon quasimomenta is clearly higher than MF, shows that the just chosen ansatz also breaks down if both the hopping and the interaction strength are too large. However, it is still possible that some aspects of the FE dispersion can be translated into real physics, e.g. that the FE predict a transition to the supersonic regime for smaller quasimomenta than MF. A remarkable feature of the MF dispersion is the kink around $q=\pm\pi/a$ at the edge of the BZ.\par
If this kink is physical, it is evident that we cannot expect to capture it with the FE using either the Fourier or the heuristic ansatz. This is attributable to the fact that the Fourier series employed to describe the FE dispersion is truncated with a hard (soft) cut-off for the Fourier ansatz (heuristic ansatz). It is not possible to describe a function around a point where the function is non-differentiable using a finite number of terms, which means that our ansatz cannot describe non-analyticities in the quasi particle dispersion.  \par
\begin{figure}
\includegraphics[width=0.48\textwidth]{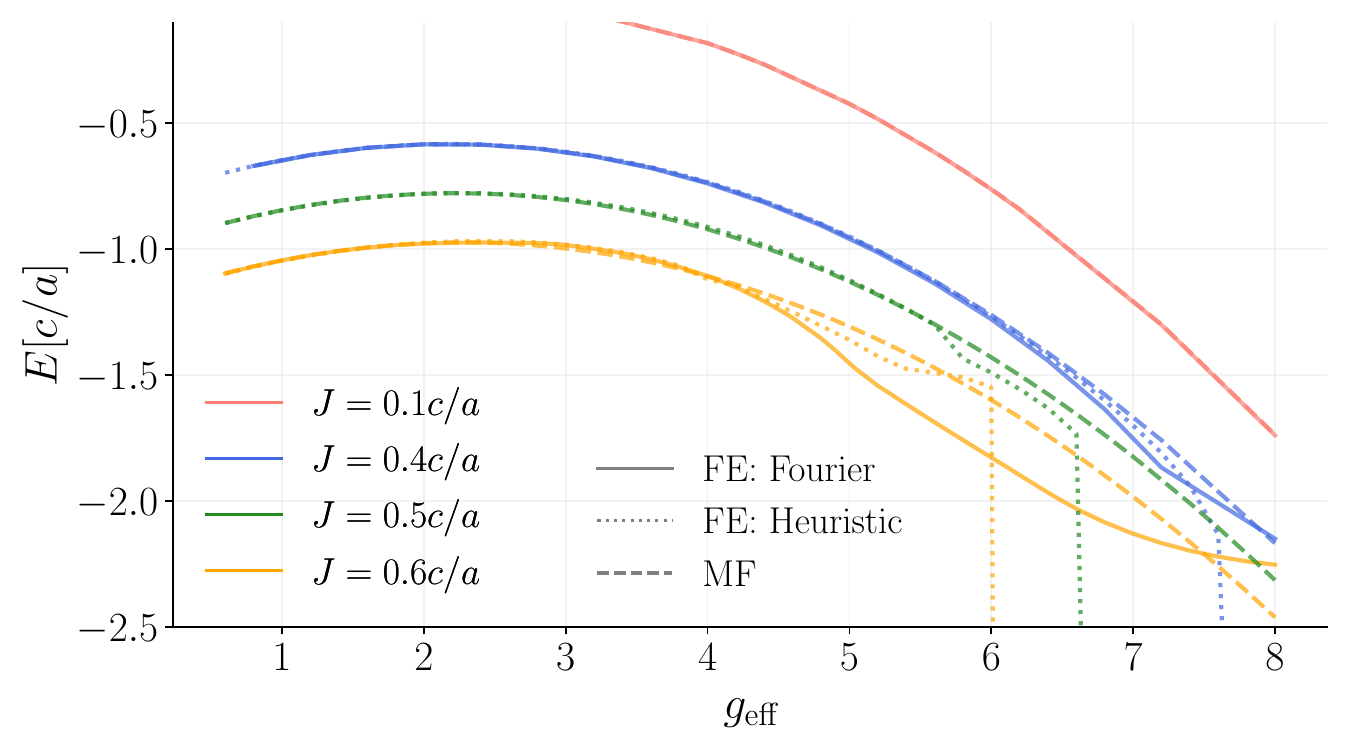}
\caption{Ground state energy at momentum $q=0$ as a function of the interaction strength for different hoppings $J$ without the inclusion of two-phonon scattering terms. We show the results according to MF (dashed lines) and the FE using a Fourier ansatz with $N=6$ (solid lines) and the heuristic ansatz (dotted lines). The colors indicate the relevant hoppings. All system parameters were chosen exactly as specified in \autoref{fig:disp_fröhlich_J=0.4}.}
\label{fig:g_scan_fröhlich}
\end{figure}

We close our discussion of the pure Fröhlich Hamiltonian by looking at the ground state energy at momentum $q=0$ as a function of the interaction strengths in \autoref{fig:g_scan_fröhlich}. For fixed hopping $J$, we observe very good agreement of the FE and MF ground state energy across a wide range of interaction strengths. This holds both for the Fourier and the heuristic ansatz. Above a certain interaction strength, which is greater for larger hoppings, the energy of the FE with Fourier ansatz cannot be considered reliable since they exceed the MF energy. This can be seen as an artifact of the oscillations already observed in \autoref{fig:disp_fröhlich_J=0.7}.  This oscillatory behavior can be reduced by using the heuristic ansatz for the FE, but eventually the ground state energy becomes unphysical for large interaction strenghts even then. In contrast to the Fourier ansatz, the breakdown of the FE is now not manifested by a decrease and subsequent increase in the ground state energy compared to MF. Instead, the ground state energy diverges to minus infinity above a critical interaction strength in a non continuous way, which is clearly not physical. \par
The analysis of the dispersion of the Fröhlich-type Hamiltonian with the flow equations using two different ansätze has shown us that the failure of the flow equations is caused in two ways:  On the one hand, if the hopping is increased for fixed $g_{\mathrm{eff}}$ we observe that the FE results may depend strongly on the FE ansatz chosen. Once relevant experiments will be conducted, conclusions could be drawn about the true dispersion and consequently also about a good ansatz for the operators subject to the flow. On the other hand, higher-order terms in the bosonic operators in the flow are generated more quickly in the FE for larger interaction strengths, which is why this approximation becomes poorer, as discussed in \autoref{sec:formalism}. However, for repulsive potentials, two-phonon scattering terms become relevant for larger interaction strengths anyway, which is why we reach the limits of the validity of the Fröhlich Hamiltonian if $|g_{\mathrm{eff}}|$ is too large. The next section is devoted to examining the effect of including the two-phonon scattering terms, leading to the extended lattice Bogoliubov Fröhlich Hamiltonian.

\subsection{The Extended Bogoliubov Fröhlich Hamiltonian}\label{sec:Results:BogFröhlich}
\subsubsection{Zero Hopping: Static Impurity}
We will start our discussion by considering the case $J=0$, i.e. the impurity is localized and cannot hop between lattice sites. This case is exactly solvable using a Bogoliubov-de Gennes transformation and shall serve as a test to assess the range of applicability of the flow equations.
\begin{figure}
\includegraphics[width=0.48\textwidth]{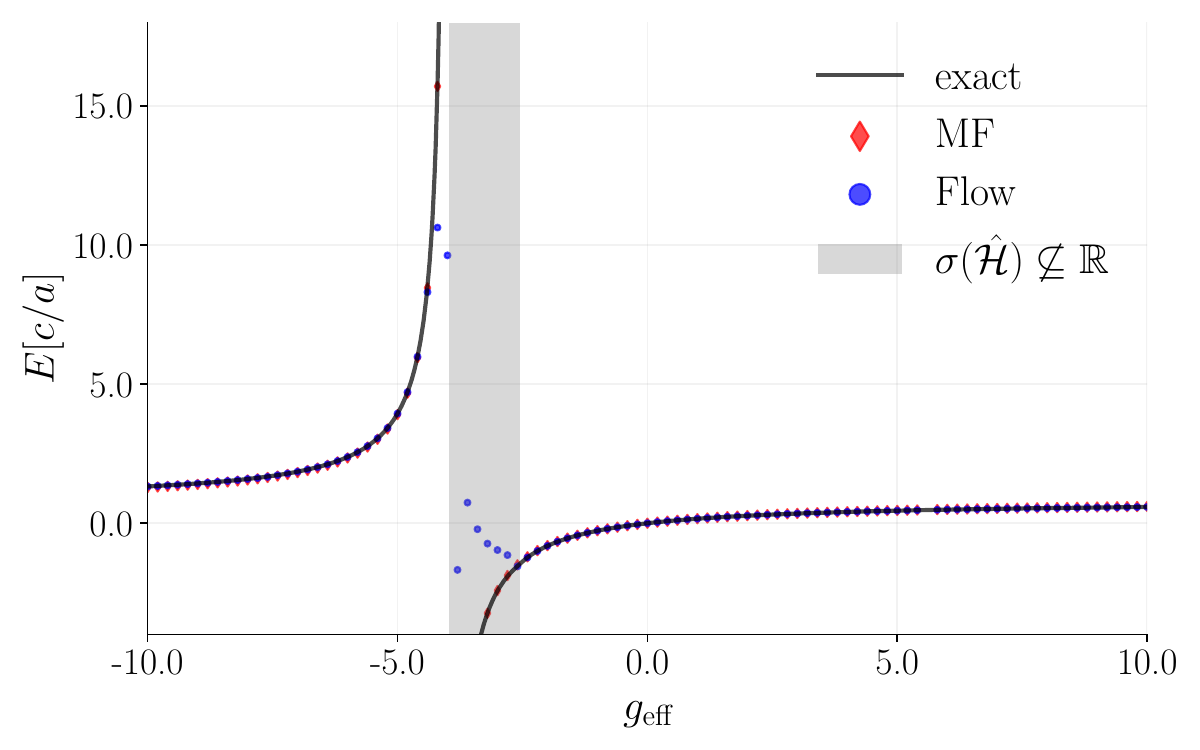}
\caption{Ground state energy as a function of the interaction strength for a localized impurity as calculated via MF (red marker), the flow equations (blue marker) and exact diagonalization (black line) with the inclusion of two-phonon scattering terms. The grey shaded area indicates the region where the system is dynamically unstable. All system parameters were chosen exactly as in \autoref{fig:disp_fröhlich_J=0.4}.}
\label{fig:g_scan_exact_comp}
\end{figure}

First of all, it is notable in \autoref{fig:g_scan_exact_comp} that above a certain critical attractive interaction strength, both MF and FE agree with the results from exact diagonalization. Below this critical interaction strength $g_{\mathrm{eff}}^c\approx -2.55$, a pair of conjugate complex eigenvalues emerges, as the exact diagonalization reveals. Then the system is said to be dynamically unstable, since the system will irreversibly fall out of its equilibrium state if perturbed infinitesimally \cite{Kain_2018}. The complex eigenvalues cannot be reproduced by the flow equations. Instead, they are encoded in such a way that not all off-diagonal terms vanish. If the interaction strength is lowered even further, the dynamical instability disappears and the FE converge again against the result from exact diagonalization. \par
At the same time, we find that the MF energy functional \eqref{eq: MF_energy_functional} is no longer bounded from below for interaction strenghts lower than $g_{\mathrm{eff}}^c$. However, there still exists a MF saddle point in good agreement with the result obtained via exact diagonalization even in this regime.\par
\begin{figure}
\includegraphics[width=0.48\textwidth]{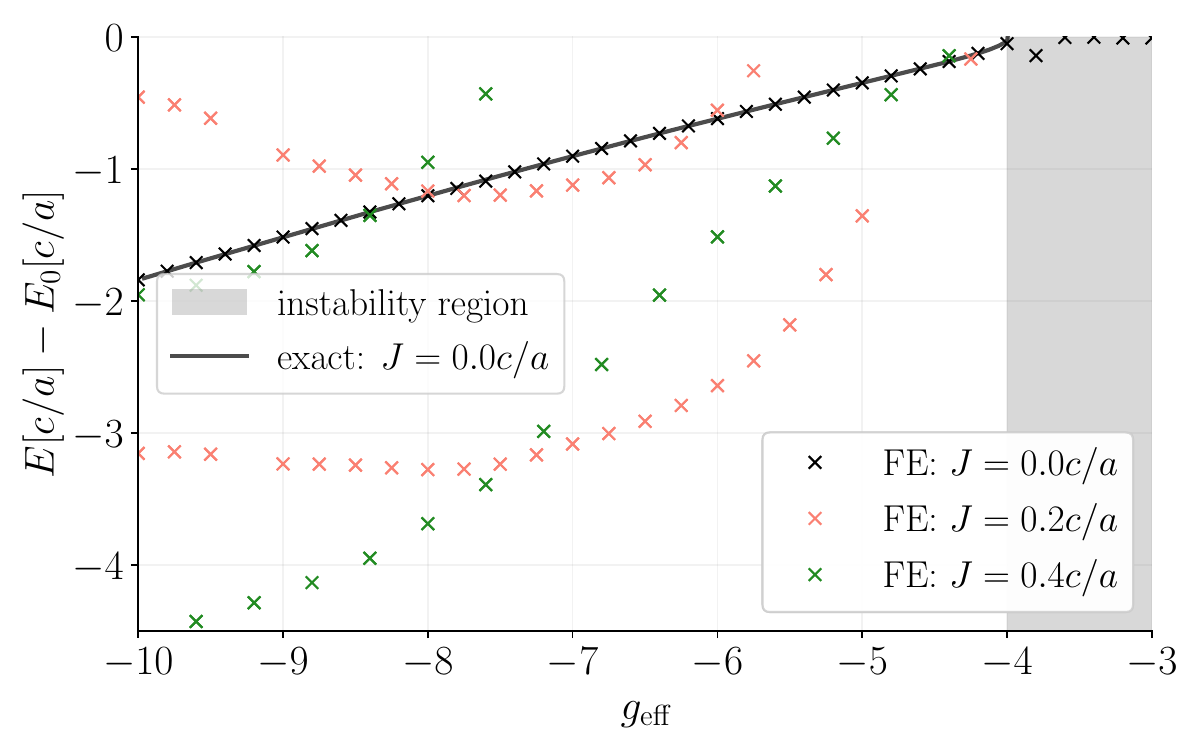}
\caption{Low energy part of the spectrum determined by the energies of single phonon Fock states $\CR_k|0\rangle$  of momentum $k$ for different interaction strengths $g_{\mathrm{eff}}$ and hoppings $J$ with the inclusion of two-phonon scattering terms. The gray shaded region indicates where an instability in the form of imaginary eigenvalues is predicted according to the exact solution for vanishing hopping $J=0.0$. The energies are with respect to the phonon vacuum $E_0$. All system parameters were chosen exactly as in \autoref{fig:disp_fröhlich_J=0.4}.}
\label{fig:g_scan_spectrum}
\end{figure}
\vspace{12pt}
Although we lack the comparitive reference of exact diagonalization in the case of nonzero hopping $J$, we still expect similar behaviour in that case: Especially near the resonance, in the regime of dynamical instability, we believe that the results from the FE are not strictly accurate. This can also be seen in \autoref{fig:g_scan_spectrum} where the flow equations predict a mode with single-particle energy $\omega=0$ in the grey-shaded region. This is in contrast to exact diagonalization, which yields a pair of imaginary, complex conjugated eigenvalues with vanishing real parts. Notably, for even lower interaction strengths, the FE correctly predict the existence of a bound state with negative single-particle energy $\omega<0$ for $J=0$. Furthermore, we see from the orange and green data in \autoref{fig:g_scan_spectrum} that the FE also predict single particle excitations for $J>0$ that energetically lie below the phonon vacuum and thus indicate the existence of a bound state.

\subsubsection{Nonzero Hopping: Mobile Impurity on a Lattice} \label{sec:results-B-2}
\begin{figure}
\includegraphics[width=0.48\textwidth]{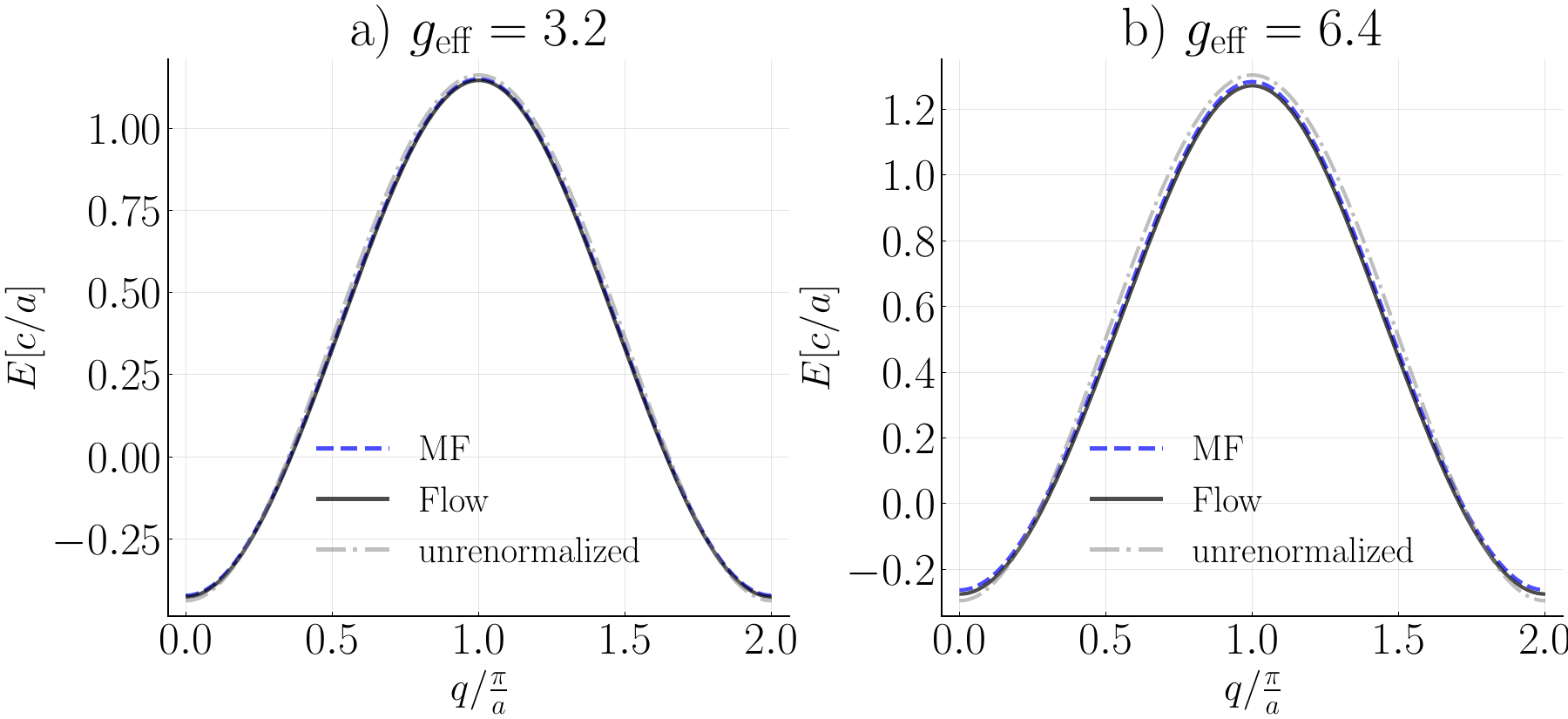}
\caption{Comparison of the polaron dispersion obtained via MF (blue line) and the flow equations (black line) with Fourier cutoff $N=6$ for fixed hopping $J=0.4c/a$ and two different positive interaction strengths $g_{\mathrm{eff}}$. The system parameters are the same as those specified for \autoref{fig:disp_fröhlich_J=0.4}. This time, however, we work with the extended Bogoliubov Fröhlich Hamiltonian which includes two-phonon-scattering terms.}
\label{fig:disp_BF_J=0.4}
\end{figure}

We open our analysis by comparing the effect of including the full BF Hamiltonian on the dispersions in \autoref{fig:disp_fröhlich_J=0.4}. In \autoref{fig:disp_BF_J=0.4}, we observe that for $g_{\mathrm{eff}}>0$ the two-phonon terms have the effect of counteracting the strong renormalization, which became particularly prominent around $q=\pm\pi/a$ at the edge of the BZ. The flow equations agree with the MF result and lie only marginally lower in terms of energy. Indeed, the fact that they have lower energies suggests that the FE are accurate in this regime.\par
\begin{figure}
\includegraphics[width=0.48\textwidth]{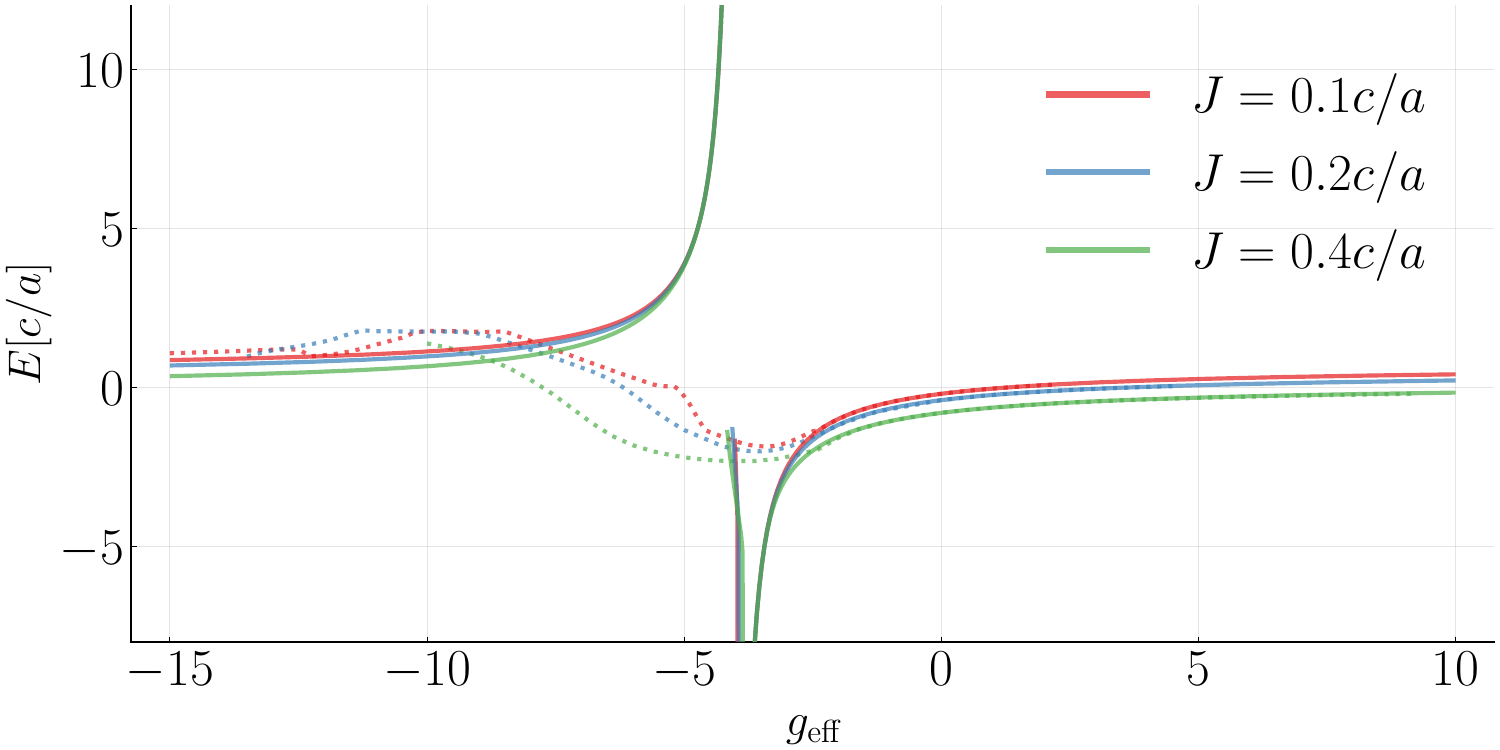}
\caption{Ground state energy at momentum $q=0$ as a function of the interaction strength for three different hoppings $J$. We show the results according to MF (solid lines) and the FE using a Fourier ansatz with $N=6$ (dotted lines). The colors indicate the relevant hoppings. All system parameters were chosen exactly as specified in \autoref{fig:disp_fröhlich_J=0.4}.}
\label{fig:g_scan_BF}
\end{figure}

The comparison with \autoref{fig:disp_fröhlich_J=0.4} also shows that for fixed $g_{\mathrm{eff}}$ due to the the inclusion of two phonon scattering terms, in addition to the change of the shape of the dispersion, the overall energy increases. 
This also manifests itself in the fact that we observe the typical Bose polaron branches in \autoref{fig:g_scan_BF}. These are known to be connected for $g_{\mathrm{eff}}\rightarrow\pm\infty$ in the case of a non-lattice Bose polaron \cite{Grusdt_2017}, which is consistent with our MF results for the lattice polaron. Within the flow equation approach, the two polaron branches are not separated by a divergence of the ground state energy but are continuously connected, see \autoref{fig:g_scan_exact_comp} and \autoref{fig:g_scan_BF}. Nevertheless, the FE also indicate that the asymptotic behavior in $g_{\mathrm{eff}}\rightarrow\pm\infty$ is connected. Furthermore, the FE very closely match and thereby confirm the MF predictions above a critical negative interaction strength $g_{\mathrm{eff}}^c\approx -2.5$. This is also the critical value below which the MF energy becomes unbounded from below. \par
\begin{figure}
\includegraphics[width=0.48\textwidth]{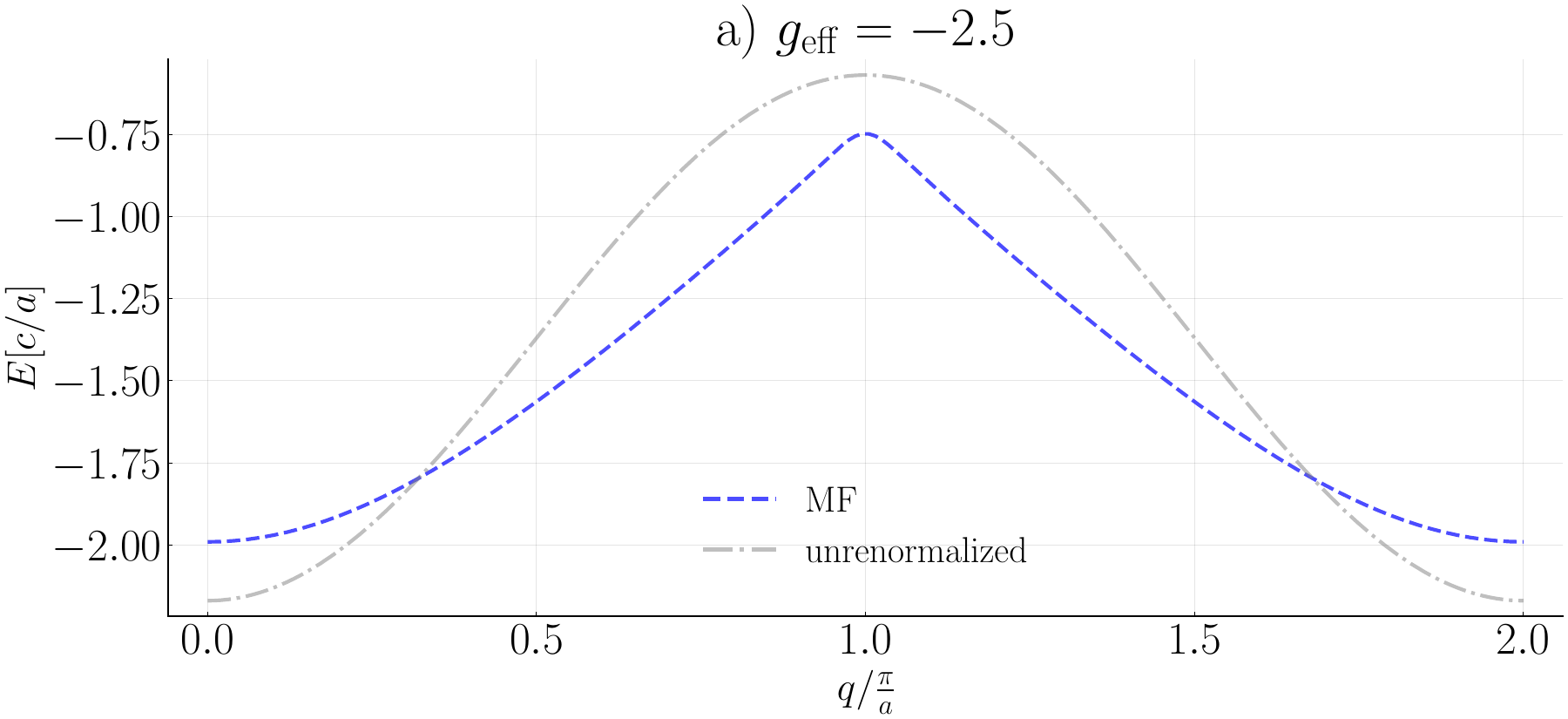}
\caption{MF dispersion for $g_{\mathrm{eff}}=-2.5$. The hopping was set to $J=0.4 c/a$. All system parameters apart from the hopping were chosen exactly as in \autoref{fig:disp_fröhlich_J=0.4}.}
\label{fig:MF_kink_two_ph}
\end{figure}

 
For large, repulsive interaction strengths, we have seen that there is no sharpening of the dispersion when including two phonon scattering terms (see \autoref{fig:disp_BF_J=0.4}).  Instead, in \autoref{fig:MF_kink_two_ph} we find such a dispersion for negative interaction strengths. Since this is in the regime where the FE are no longer accurate due to the emergence of strong oscillatory behavior, we have not included the associated FE dispersion in this plot.\par 

\section{Formalism}
\label{sec:formalism}
\subsection{Overview of the Flow Equation Approach} \label{sec:formalism:overview}
The flow equation approach was introduced by \citet{WegnerOG_Article} and by \citet{WilsonOG_Article} as a new renormalization technique to diagonalize or block-diagonalize Hamiltonians. It involves applying a continuous unitary transformation, guided by an appropriately chosen generator, to the Hamiltonian, which increasingly diagonalizes the Hamiltonian.\par
Generically, the starting point is a Hamiltonian $\ham$, which we can split into a diagonal part $\ham_0$ and an off-diagonal/ interaction part $\ham_{\mathrm{int}}$ such that
\begin{equation}
\ham = \ham_0+\ham_{\mathrm{int}}.
\end{equation} 
Applying a continuous unitary transformation $\hat U(\lambda)$ ordered by the flow parameter $\lambda$ with $[\lambda ]=\mathrm{Energy}^{-2}$ is intended to transform this Hamiltonian into a unitarily equivalent diagonal Hamiltonian, i.e. the off-diagonal elements in the flow Hamiltonian \begin{equation}\label{def: flow ham} \ham(\lambda)=\hat U(\lambda)\ham(\lambda=0)\hat U^\dagger(\lambda)\end{equation} 
should vanish in the limit $\lambda\rightarrow\infty$. $\hat U(\lambda)$ is connected to the antihermitian flow generator $\hat\eta$ by 
\begin{equation}\label{def: generator}
\hat\eta(\lambda)\defeq \frac{\mathrm d \hat U(\lambda)}{\mathrm d\lambda} \hat U^\dagger(\lambda)=-\hat\eta^\dagger(\lambda).
\end{equation}
Differentiating eq. \eqref{def: flow ham} w.r.t. $\lambda$ and inserting eq. \eqref{def: generator} gives the following differential equation which we shall refer to as the flow equations for the Hamiltonian $\ham(\lambda)$,
\begin{equation}\label{def: flow equation}
\frac{\mathrm d\ham(\lambda)}{\mathrm d\lambda}=\left[\hat\eta(\lambda),\ham(\lambda) \right].
\end{equation}
From now on, the $\lambda$-dependence of $\hat\eta,\ham$ and $\hat U$ will be notationally dropped.\par
Throughout the present article, we will work with
\begin{equation}
\hat\eta = \left[\ham_0,\ham_{\mathrm{int}}\right].
\end{equation}
This choice is called the canonical generator and usually achieves the desired diagonalization, as discussed in Ref. \cite{kehrein2006flow}. It is also elaborated there that the fundamental difficulty of the flow equation approach is that, depending on the structure of the Hamiltonian, the flow equation \eqref{def: flow equation} may generate new, higher-order interaction terms during the flow. We will encounter precisely this problem with the lattice Bose polaron. In the next section, we will discuss the truncation scheme that is necessary to handle this proliferation of terms.\par

\subsection{Transitioning the Flow Parameters to Operators} \label{sec:formalism:transition}
The flow equation approach can be successfully applied to the free Bose polaron in the heavy impurity limit when employed in combination with a coherent state ansatz. For this problem, the flow equations are able to reproduce in excellent agreement the spectrum obtained from exact diagonalization by means of a coherent state ansatz and a Bogoliubov-de Gennes transformation. And indeed, in \autoref{sec:results} we have seen that except in the vicinity of the resonance, the flow equations are able to reproduce the ground state energy obtained via exact diagonalization.\par
For $J\neq 0$, the Hamiltonian \eqref{eq: ham_q_with_2phon} can of course still be split into a diagonal and off-diagonal part, but the diagonal part now depends on the bosonic momentum operator $\hat P$ which is diagonal in the occupation number representation. It is clear that this dependence on the momentum operator, whose strength is controlled by the hopping $J$, will carry over into the off-diagonal part during the flow. This is further supported by the \emph{formal} similarities to the work of \citet{PhysRevA.106.033321}, as we will discuss in \autoref{sec:formalism:breakdown}. This motivates choosing a quadratic Hamiltonian as an ansatz for the flow Hamiltonian, where the coefficients are not $c$-numbers but operators which are diagonal in the occupation number representation $\hat n_k= \CR_k\AN_k$. \par
It is convenient to express the flowing Hamiltonian in the following manifestly hermitian form:
\begin{widetext}
\begin{align}
\ham & =\ham_0+\ham_{\mathrm{int}}=\ham_0+\Sum_{q\neq q^\prime}\hat V_{q,q^\prime} \CR_q\AN_{q^\prime}+\Sum_{p,p^\prime}\left(\hat W_{p,p^\prime} \CR_p\CR_{p^\prime}+\AN_p\AN_{p^\prime}\hat W_{p,p^\prime}\right)+\Sum_k \left(\hat U_k\CR_k+\AN_k\hat U_k\right), \label{gen_ham}
\end{align}
\end{widetext}
where all new operators $\hat V,\hat W,\hat U$ that have been introduced functionally depend on the number operators $\{\hat n_k\}_k$ with $\hat n_k\defeq \CR_k\AN_k$. It is clear that the lattice polaron Hamiltonian \eqref{eq: ham_q_with_2phon} is of this form, after the integrals have been discretized into sums and the occurring prefactors have been absorbed into the relevant parameters.
Careful inspection of \begin{equation} \label{eq:diag_part_def}
\ham_0\eqdef \hat H_0+\Sum_k \hat H_k \CR_k\AN_k,
\end{equation}  and $\ham_{\mathrm{int}}$ reveals that the canonical generator will contain normal-ordered terms with three or more creation or annihilation operators. This means that during the flow, the Hamiltonian \eqref{gen_ham} would acquire even higher order terms which we will trunate at second order. Whenever such a truncation has been performed, we shall indicate it with the symbol "$\eqsec$" and refer to it as the quadratic approximation.\par
Since we are working with normal ordered expressions, the quadratic terms contain all the information about the one-particle energies, the quartic terms contain the information about two-particle interactions and so forth. In our case, we perform the normal ordering with respect to the vacuum, since in the subsonic regime of the lattice polaron it can be assumed that the bosonic vacuum is indeed the ground state of the system. Therefore it is expected that this truncation is well behaved in our case. For other physical problems, it may be necessary to adapt our formalism to allow for normal ordering with respect to states other than the vacuum \cite[sect. 4.1.5]{kehrein2006flow}.\par
Within this approximation, the canonical generator $\hat\eta$ is given by
\begin{align}
\hat\eta &\defeq \left[\ham_0,\ham_{\mathrm{int}}\right] \\
&\eqsec  \Sum_{q\neq q^\prime}\hat \theta_{q,q^\prime} \CR_q\AN_{q^\prime}+\left(\Sum_{p,p^\prime}\hat \phi_{p,p^\prime} \CR_p\CR_{p^\prime}+\Sum_k  \hat\varsigma_k \CR_k-\mathrm{h.c.}\right).\label{def: eta_via_theta_phi_sigma}
\end{align}
Here we introduced both the auxiliary operators
\begin{align}
\hat\theta_{q,q^\prime}&\defeq\hat V_{q,q^\prime} \Big(\hat H_0 - \hat H_0(\hat n_q-1,\hat n_{q^\prime}+1)\nonumber\\ &+\hat H_{q} - \hat H_{q^\prime}(\hat n_q-1,\hat n_{q^\prime}+1)\Big),\label{def: theta}\\
\hat\phi_{p,p^\prime}&\defeq\hat W_{p,p^\prime}\Big(\hat H_0-\hat H_0(\hat n_{p^\prime}-1,\hat n_p -1)\nonumber\\ &+\hat H_p+\hat H_{p^\prime}\Big), \label{def: phi}\\
\hat\varsigma_k&\defeq\hat U_k\left(\hat H_0-\hat H_0(\hat n_k -1)+\hat H_{k}\right), \label{def: sigma}
\end{align}
and the notation $\hat O(\hat n_k\pm 1)$, where $\hat O$ is some operator diagonal in the occupation number representation. We define $\hat O(\hat n_k\pm 1)$ implicitly by
\begin{align}
\AN_k\hat O(\hat n_k-1)&= \hat O\AN_k, \label{def: n-not1}\\
\CR_k\hat O(\hat n_k+1)&= \hat O\CR_k,\label{def: n-not2}
\end{align}
and we extend our notation in the natural way to allow for two arguments:
\begin{align}
\hat O(\hat n_k\pm1,\hat n_{k^\prime}\pm1)\defeq \left(\hat O(\hat n_k\pm1)\right)(\hat n_{k^\prime}\pm1).
\end{align}
This is best understood when considering the action of $\hat O(\hat n_k\pm 1)$ on a Fock state $|\nu\rangle\defeq|\tilde n_1, \tilde n_2,\ldots\rangle$ with occupation number $\tilde n_i$ for state $i$. By assumption, this is an eigenstate of $\hat O$:
\begin{align}
\hat O|\tilde n_1, \tilde n_2,\ldots\rangle\eqdef O_{\tilde n_1, \tilde n_2,\ldots}|\tilde n_1, \tilde n_2,\ldots\rangle.
\end{align}
It then follows from eqs. \eqref{def: n-not1} and \eqref{def: n-not2} that
\begin{align}
\hat O(\hat n_k\pm1)|\nu\rangle&= O_{\tilde n_1, \tilde n_2,\ldots,\tilde n_k\pm1,\ldots}|\nu\rangle.
\end{align}
However, the action of $\hat O(\hat n_k-1)$ on states with $\tilde n_k=0$ is ill-defined, since the annihilation operator $\AN_k$ in eq. \eqref{def: n-not1} is invertible only on subspaces consisting exclusively of states with $\tilde n_k>0$. For our purposes, though, this is not much of a concern, since operators of the type \eqref{def: n-not1} and \eqref{def: n-not2} enter the flow equations via the following commutators:
\begin{align}
\left[\hat O,\CR_k\right]&=\left(\hat O-\hat O(\hat n_k-1)\right)\CR_k ,\label{eq: comm_a_dagger}\\
\left[\hat O,\AN_k\right]&=\left(\hat O-\hat O(\hat n_k+1)\right)\label{eq: comm_a}\AN_k.
\end{align}
It is now clear that $\hat O(\hat n_k-1)$ only ever acts on states where site $k$ is occupied, hence it suffices for its action to be defined only on such states.\par
Eq. \eqref{def: eta_via_theta_phi_sigma} is derived essentially from the repeated use of the two identities \eqref{eq: comm_a_dagger} and \eqref{eq: comm_a}. \par 
To arrive at the flow equations, the commutator of the Hamiltonian \eqref{gen_ham} and the canonical generator \eqref{def: eta_via_theta_phi_sigma} has to be evaluated. For the commutator $\ham_0$ and $\hat\eta$ no extra work is required since $\hat \eta$, which itself is the commutator of $\ham_{\mathrm{int}}$ and $\ham_0$, is structurally identical to $\ham_{\mathrm{int}}$ within the quadratic approximation. After repeated use of \eqref{eq: comm_a_dagger} and \eqref{eq: comm_a} and the canonical bosonic commutation relations, we arrive at a closed set of flow equations valid within the quadratic approximation as discussed above. For the explicit form of the flow equations we refer to Appendix \ref{sec:app-A}.\par
\subsection{Choosing a suitable Ansatz for the Operators} \label{sec:formalism:ansatz}
Since the flow equations we have derived are operator-valued, it is now necessary to choose a well-behaved ansatz for the operators that allows the differential equations to be solved numerically. First, we will choose a generic Fourier series type ansatz, which may produce unreasonable results in certain parameter regimes. Based on the phenomenological observation that mean field theory and the flow equations quantitatively agree when the Fourier expansion coefficients associated with higher energies are strictly smaller than those associated with lower energies, in a second step we will choose a heuristic ansatz that enforces precisely that condition.
\subsubsection{Fourier Series Ansatz} 
The fact that initially the only dependence on the number operators lies in the cosine of the boson momentum is a great simplification, i.e. all we have is a $\cos\left(aq-a\hat P\right)$-type dependence. The flow will also create higher frequency terms of the form $\cos\left(naq- n a\hat P\right),\ n\in\N$. This is because in the first integration step the terms created by the commutator from the canonical generator and the diagonal part will clearly generate $\cos^2\left(aq-a\hat P\right)$-type terms, which
by trigononmetric identities contain a component with twice the frequency. In the subsequent step, the flow Hamiltonian will also pick up
these higher frequency term, which will in turn create even higher frequency terms. This
process of generating higher frequency terms makes it clear that the contribution of a term $\cos\left(naq-n a\hat P\right)$ should be of no order lower than $J^n$ because higher frequency terms are created by multiplication of lower frequency terms. And indeed, the calculations leading to eq. \eqref{eq: OP2} demonstrate that the contribution of a term $\cos\left(naq-n a\hat P\right)$ is of order $J^n$, although contributions $J^m,m>n$ are also possible due to the flowing back from higher terms of the Fourier series. For sufficiently small $J$, however, these contributions should be negligible. \par 
Furthermore, the operators in definitions \eqref{def: theta}-\eqref{def: sigma} will introduce sine terms, as eq. \eqref{eq: OP1} shows. Hence a suitable ansatz for an operator $\hat O$ in the flow Hamiltonian is given by a Fourier series:
\begin{align} \label{eq: Fourier_ansatz}
\hat O = \alpha_0^{(O)}+&\Sum_{n=1}^\infty \Big[\alpha_n^{(O)}\cos\left(naq-n a\hat P\right)\nonumber\\ &\quad\ +\beta_n^{(O)}\sin\left(naq-n a\hat P\right)\Big].
\end{align}
In practice, we are only able to consider partial sums thereof and thus have to truncate the Fourier series at order $N\in\N$, allowing us to represent each operator by $2N+1$ $c$-numbers. One advantage of this ansatz is that it requires minimal effort to deal with arbitrary optical potentials to which the impurity is coupled, provided that they can be approximated sufficiently well by a Fourier series (and do not violate the tight-binding approximation \eqref{eq: tight-binding_ham}). This allows for the adaptation to any experimental deviations from the idealized potential investigated here. \par In order to approach solving our flow equations numerically, we need to understand both how products of these operators and operations such as \eqref{def: n-not1} or \eqref{def: n-not2} can be evaluated. For this, please refer to Appendix \ref{sec:app-B}.\par
In practice, it is easy to implement the representation of the operators in such a way that $N$ can be arbitrarily chosen, tuning the accuracy of the approximation. In contrast, the ansatz of an at most quadratic flow Hamiltonian is rather fixed. The validity of this approximation is primarily determined by the interaction strength (in our case $g_{\mathrm{IB}}$), as this governs how quickly higher-order terms are generated in the flow.\par
\subsubsection{Heuristic Ansatz} \label{sec:heuristic_ansatz}
There may be cases where the functional dependence of the operators, especially the diagonal part $\hat H_0,\hat H_k$, cannot be captured by a truncated Fourier series. Then, the truncation scheme just developed, where terms of order $N'>N$ are merely ignored, need not be well-behaved, because the infinite number of terms beyond the cutoff may have made a finite contribution to one or more of the $2N+1$ expansion parameters.\\
However, the operator dependence of the diagonal part can be translated quite directly into the shape of the dispersion, so that we can find a reasonable operator ansatz based on our predictions of the shape of the dispersion. A suitable heuristic ansatz for the bosonic lattice polaron should fulfill the following conditions:
\begin{itemize}
\item The Fourier spectrum of the ansatz should fall off sufficiently quickly, as we expect the dispersion to qualitatively resemble a cosine curve: At least for small hoppings $J$, we expect only a single local and global maximum of the dispersion, located at the edge of the BZ at $q=\pm\pi/a$.
\item \autoref{fig:g_scan_exact_comp} suggests that for large effective interaction strengths, MF should essentially be correct. A suitable ansatz should therefore be able to capture the narrowing of the dispersion at the edge of the BZ, as seen in  \autoref{fig:disp_fröhlich_J=0.4}.
\item It should be possible to represent our ansatz by a manageable number of basis functions. These should be closed in the sense that products of them allow themselves to be expressed at least approximately by a linear combination of the same basis functions. The same should hold for operations \eqref{def: n-not1} and \eqref{def: n-not2}.
\end{itemize}
One possible choice is given by the following set of functions:
\begin{align} \label{eq:heur_ansatz_basis}
c_i(x) \defeq -1/2 &+ \frac{\sinh(i)}{\cosh(i)-\cos(x)}, \\
s_i(x) \defeq &-\frac{\sin(x)/2}{\cos(x)-\cosh(i)}.
\end{align}
For these functions to be continuous, we have to require $i>0$, e.g. $i\in\{1,2,3,4\}$. In that case, their Fourier spectrum has the convenient property that higher frequency contributions are exponentially (but smoothly) suppressed, thus enforcing the phenomenological condition $\alpha_{n+1}^{(H_0)}<\alpha_{n}^{(H_0)}\ \forall 0< n<N$, we have found in \autoref{sec:results}. We do not expect these functions to approximate any function defined on the BZ, but we do expect them to be able to capture the dispersion sufficiently well.\par
In a similar form to how we have represented the operators by a truncated Fourier series in eq. \eqref{eq: Fourier_ansatz}, we use the basis functions to write any operator in the flowing Hamiltonian as
\begin{equation} \label{eq: heur_ansatz_operator}
\hat O = \tilde\alpha_0^{(O)}+\Sum_{i=1}^{\tilde N} \Big[\tilde\alpha_i^{(O)}c_i\left(aq- a\hat P\right)+\tilde\beta_i^{(O)}s_i\left(aq-a\hat P\right)\Big],
\end{equation}
where $\tilde N\in\N$ is the truncation cutoff.
\subsection{Solving the Flow Equations}\label{sec:formalism:solving}
In combination with a suitable ansatz for the operators, as discussed in the previous section, the operator-valued flow equations in Appendix \ref{sec:app-A} effectively turn into flow equations for the expansion coefficients. While it would be possible to simply substitute the relevant ansätze for the operators into the flow equations and derive explicit flow equations for the coefficients, we instead choose the following procedure: Within the Fourier ansatz (heuristic ansatz), every operator is uniquely determined by $2N+1$ ($2\tilde N+1$) $c$-numbers. This allows for each operator to be represented by an array of $2N+1$ (or $2\tilde N+1$) $c$-numbers. For the Fourier ansatz, we define a multiplication of two such arrays according to the calculations shown in Appendix \ref{sec:app-B}. For the heuristic ansatz, we calculate how the products of the basis functions can be expanded in the basis, which only needs to be done once and therefore does not add to the computational cost. In this way, we generate a multiplication for the arrays of size $2\tilde N+1$.
In a similar spirit, we evaluate how the arrays transform under operations of the form $\hat O(\hat n_q\pm1,\hat n_{q^\prime}\pm1)$. For the Fourier ansatz, the relevant calculations are shown in Appendix \ref{sec:app-B}.\par
Equipped with these multiplication and transformation rules, we can numerically solve the flow equations in Appendix \ref{sec:app-A} at the operator level. We get a set of coupled, non-linear ordinary differential equations (ODE) that are easy to solve in principle. The challenge lies in the number of ODEs we have to solve: We need $(2N+1)(N_{\mathrm{grd}}+1)$ operators to describe the diagonal part $\ham_0$ of the flowing Hamiltonian \eqref{gen_ham}. The interaction part $\ham_{\mathrm{int}}$ requires another $2N_{\mathrm{grd}}^2(2N+1)$ parameters, which means that we are dealing with a set of $(2N+1)(2N_{\mathrm{grd}}^2+N_{\mathrm{grd}}+1)$ coupled ODEs. This number scales unfavorably with the system size $N_{\mathrm{grd}}$.\par
Having traversed the flow, i.e. solved the ODEs, for a sufficiently long intervall in the flow parameter $\lambda$, the interaction part of the Hamiltonian is sufficiently suppressed and we can assume that the diagonal part now contains all the information about the dispersion and the ground state energy that we are interested in. If no bound state is present, the ground state of the flowing Hamiltonian turns out to be the phonon vacuum where the total bosonic momentum $\hat P=0$ vanishes. A priori, this is not clear from the form of $\ham_0$ and has to be verified numerically for each quasimomentum $q\in\mathrm{BZ}$.\par
If we were interested in observables other than energy, e.g. the phonon density around the impurity, results could not be found that straightforwardly. For a general observable $\hat O$, we need to apply the same series of unitary transformations that have been applied to the Hamiltionian. This is equivalent to solving another set of ODEs \cite{kehrein2006flow}:
\begin{equation}\label{def: flow equation_observables}
\frac{\mathrm d\hat O(\lambda)}{\mathrm d\lambda}=\left[\hat\eta(\lambda),\hat O(\lambda) \right].
\end{equation}

\subsection{Breakdown of the Flow Equations}\label{sec:formalism:breakdown}
In the following, we will try to give an argument on why the flow equations break down in the area where MF predicts a divergence, which we have discussed in \autoref{sec:Results:BogFröhlich}. \\
Specifically, \citet{PhysRevA.106.033321} has demonstrated for the Bose polaron with quadratic dispersion that an impurity can be analytically described as a deformation in the Lie algebra defined by the bosonic operators. This deformation manifests itself (for a single bosonic mode) in modified commutation relations for the deformed bosonic operators $\hat b,\hat b^\dagger,\hat B$:
\begin{equation}
\hat b\hat b^\dagger - \zeta\hat b^\dagger\hat b = 1,\ \left[\hat B,\hat b^\dagger\right] = \hat b^\dagger,\ \left[\hat B,\hat b\right] = - \hat b,
\end{equation}
where $\zeta$ is called the deformation parameter. The deformed operator $\hat b$ can be expressed in terms of the original bosonic operators in the form of an infinite series
\begin{equation} \label{eq: deformation_series}
\hat b^\dagger = \Sum_{n=0} c_n (\CR\AN)^n \CR,
\end{equation}
with the expansion coefficients $c_n$ scaling with $\zeta$. The original polaron Hamiltonian in the LLP frame can then be understood as a series expansion of uncoupled but deformed harmonic oscillators. The corresponding Hamiltonian can be organized in the form of eq. \eqref{gen_ham}. \par
Crucially, \citet{PhysRevA.106.033321} has shown that at the transition from the repulsive to the attractive polaron (where the divergence of the ground state energy can be seen), the deformation parameter diverges as well. However, if the deformation parameter diverges, any truncation of the series \eqref{eq: deformation_series} is no longer valid. Translated to our flow equations, this means that a finite representation of the expansion operators can no longer be justified and the approximation discussed in \autoref{sec:formalism:ansatz} cannot be applied.\par
These considerations are valid for the Bose polaron with a quadratic dispersion, but while the analysis by \citet{PhysRevA.106.033321} cannot be straightforwardly applied to our problem, it should be at least qualitatively applicable to the lattice Bose polaron, where MF also predicts a repulsive-attractive transition (see \autoref{fig:g_scan_BF}). After all, the cosine can be approximated by a quadratic function, at least in those Hilbert subspaces in which the total boson momentum is small.\par
This suggests that it is still useful to think about the lattice polaron in terms of boson deformations caused by the impurity. If the deformations diverge, deformed bosons cannot be considered good degrees of freedom and the \emph{formal} similarities to our method plausibilize the breakdown of the flow equations.

\section{Summary and Outlook}
\label{sec:summary}
We investigated a system where a single impurity is confined to a one-dimensional optical lattice and becomes dressed by Bogoliubov phonons of a quasi-1D BEC. We analyzed how the ground state energy changes if the theoretical model describing that system is extended to include two-phonon scattering events \cite{PhysRevA.88.053632}, calculated the renormalized dispersion and considered the ground state energy for fixed quasimomentum $q$ both via a mean-field treatment and a flow equation approach. \par
We have found that the strong renormalization of the dispersion, which was previously predicted by \citet{Grusdt_2014_Bloch} when treating the lattice polaron within the Fröhlich paradigm at large interaction strengths, does not prevail on the repulsive side of the Feshbach resonance after taking two-phonon processes into account. Instead, we found the same to arise at negative interaction strenghts on the attractive side of the resonance.\par
Furthermore, looking at the ground state energy for vanishing quasimomentum using mean-field theory, we were able to find the typical polaron branches, whose qualitative features remain intact for different non-zero hoppings.\par 
On the more technical side, we have introduced an extension of the flow equation approach in this article.
Formally, it involves the derivation of flow equations for a bosonic quadratic Hamiltonian where the coefficients are promoted to diagonal operators in the occupation number representation. We were able to expand these operators in a Fourier series that is dependent on the total bosonic momentum.\par
We then applied this method to the lattice polaron and showed that if the flow equations produce physical results, i.e. if no parameters diverge during the flow, they essentially confirm the results obtained via a mean field treatment: When no processes beyond the Fröhlich paradigm are considered, the flow equations are well applicable for not too large $g_{\mathrm{eff}}$. And if two phonon processes are considered, MF and the flow equations essentially agree above a (negative) impurity-boson interaction strength. However, in the transition region between repulsive and attractive polaron, the flow equations do not give physical results. We argued that this is due to the fact that in this region the operators which are subject to the flow cannot be represented by a finite number of parameters. Despite this, the flow equations suggest, as does MF, that the asymptotic behavior of the ground state energy for $g_{\mathrm{IB}}\rightarrow\pm\infty$ is identical.\par
We must conclude that the practical value of the extension of the flow equation approach developed here is rather limited for the bosonic lattice polaron. One of the key challenges involved is finding a suitable representation of the operators that are subject to the flow. Of course, this is highly dependent on the physical system under consideration, which is why it is at this point that we would like to stress that the Fourier series treatment is very specific to the lattice polaron and the form of the potential to which the impurity is coupled. \par
However, similar techniques may be applied to other problems. This is especially true for problems where initially only $\hat H_0$ depends on the number operators. In this case, it suffices to find an ansatz that is closed under multiplication and operations \eqref{def: n-not1}, \eqref{def: n-not2}. \par
For example, in the case of a free polaron with a quadratic dispersion, a polynomial ansatz instead of the Fourier series could be chosen. Here, too, a cutoff $N$ must be introduced. This cutoff would correspond to the degree of the polynomial. But again, the flow equations would only be valid within the quadratic approximation, since higher-order terms in the boson operators will also be generated  in the flow.

\section{Acknowledgments}
\label{sec:acknowledgments}
This research was funded by the Deutsche Forschungsgemeinschaft (DFG, German Research Foundation) under Germany's Excellence Strategy -- EXC-2111 390814868 -- and has received funding from the European Research Council (ERC) under the European Union’s Horizon 2020 research and innovation programm (Grant Agreement No. 948141) - ERC Starting Grant SimUcQuam.

\appendix

\onecolumngrid 
\section{The Flow Equations}
\label{sec:app-A}
The flow equations for the diagonal part of the Hamiltonian are:
\begin{subequations}
\begin{align}
\partial_\lambda\hat H_0&\eqsec -2\Sum_{p, p^\prime}\hat \phi^\prime_{p,p^\prime} \left(\hat W_{ p, p^\prime}(\hat n_p+1,\hat n_{p^\prime}+1)+\hat W_{ p^\prime,p}(\hat n_p+1,\hat n_{p^\prime}+1)\right) \\
&\ \ \ -2\Sum_{k} \hat\varsigma^\prime_k \hat U_{k}(\hat n_k+1),\nonumber\\	
\nonumber
\partial_\lambda\hat H_k&\eqsec \Sum_{p} \hat \theta_{k,p}\hat V_{p,k}(\hat n_{p}+1,\hat n_k-1) \\		
&-\Sum_{p} \hat V_{k,p}\hat\theta_{p,k}(\hat n_{p}+1,\hat n_k-1)\nonumber\\	
&-2\Sum_{q}\left(\hat \phi^\prime_{k,q}+\hat \phi^\prime_{q,k}\right) \left(\hat W_{q,k}(\hat n_{k}+1,\hat n_{q}+1)+\hat W_{k, q}(\hat n_{k}+1,\hat n_{q}+1)\right)\nonumber\\	
&+2\left(\hat U_{k}\hat\varsigma^\prime_k(\hat n_{k}-1)-\hat\varsigma^\prime_k \hat U_{k}(\hat n_k+1)\right)\nonumber.\\	
&\nonumber	
\end{align}
\end{subequations}
The off-diagonal operators follow the following equations:
\begin{subequations}
\begin{align}
\partial_\lambda\hat V_{q,q^\prime}&\eqsec -\hat V_{q,q^\prime} \left(\hat H_0 - \hat H_0(\hat n_q-1,\hat n_{q^\prime}+1)+\hat H_{q} - \hat H_{q^\prime}(\hat n_q-1,\hat n_{q^\prime}+1)\right)^2  \\	
&-\Sum_{p} \hat V_{q,p}\hat\theta_{p,q^\prime}(\hat n_{p}+1,\hat n_q-1) \nonumber\\	
&+\Sum_{p} \hat \theta_{q,p}\hat V_{p,q^\prime}(\hat n_{p}+1,\hat n_q-1)\nonumber\\	
&-\Sum_{k}\left(\hat \phi^\prime_{q^\prime,k}+\hat \phi^\prime_{k,q^\prime}\right) \left(\hat W_{k,q}(\hat n_{q^\prime}+1,\hat n_{k}+1)+\hat W_{q, k}(\hat n_{q^\prime}+1,\hat n_{k}+1)\right)\nonumber\\	
&-\Sum_{k}\left(\hat \phi^\prime_{q,k}+\hat \phi^\prime_{k,q}\right) \left(\hat W_{k,q^\prime}(\hat n_{q}+1,\hat n_{k}+1)+\hat W_{q^\prime, k}(\hat n_{q}+1,\hat n_{k}+1)\right)\nonumber\\	
&+\left(\hat \varsigma_{q}\hat U_{q^\prime}(\hat n_{q}-1,\hat n_{q^\prime}+1)-\hat U_{q^\prime}(\hat n_{q^\prime}+1) \hat \varsigma_{q}(\hat n_{q^\prime}+1)\right)\nonumber\\	
&+\left(\hat U_{q}\hat\varsigma^\prime_{q^\prime}(\hat n_{q}-1)-\hat\varsigma^\prime_{q^\prime} \hat U_{q}(\hat n_{q^\prime}+1)\right),\nonumber\\
\partial_\lambda\hat W_{p,p^\prime}&\eqsec -\hat W_{p,p^\prime}\left(\hat H_0-\hat H_0(\hat n_{p^\prime}-1,\hat n_p -1)+\hat H_p+\hat H_{p^\prime}\right)^2\\	
&+\Sum_{ q} \hat\theta_{ p,q}\left(\hat W_{q,p^\prime}(\hat n_{q}+1,\hat n_p-1)+\hat W_{p^\prime,q}(\hat n_{q}+1,\hat n_{p}-1)\right)\nonumber\\	
&-\Sum_{ q} \hat V_{ p,q}\left(\hat \phi_{q,p^\prime}(\hat n_{q}+1,\hat n_p-1)+\hat \phi_{p^\prime,q}(\hat n_{q}+1,\hat n_p-1)\right)\nonumber\\	
&+ \hat\varsigma_p\hat U_{p^\prime}(\hat n_p-1)-\hat U_{p^\prime}\hat\varsigma_p(\hat n_{p^\prime}-1),\nonumber\\						
\partial_\lambda\hat U_k&\eqsec -\hat U_k\left(\hat H_0-\hat H_0(\hat n_k -1)+\hat H_{k}\right)^2\\
&+\Sum_{q\neq k} \hat\theta_{k,q}\hat U_q(\hat n_{k}-1,\hat n_{q}+1)\nonumber\\	
&-\Sum_{\tilde p} \hat U_{\tilde p}(\hat n_{\tilde p}+1)\left(\hat \phi_{\tilde p,k}(\hat n_{\tilde p}+1)+\hat \phi_{k,\tilde p}(\hat n_{\tilde p}+1)\right)\nonumber\\	
&-\Sum_{q\neq k} \hat V_{k,q}\hat \varsigma_q(\hat n_{k}-1,\hat n_{q}+1)\nonumber\\	
&-\Sum_{\tilde p} \hat \varsigma^\prime_{\tilde p}\left(\hat W_{\tilde p,k}(\hat n_{\tilde p}+1)+\hat W_{k,\tilde p}(\hat n_{\tilde p}+1)\right).\nonumber
\end{align}
\end{subequations}
The auxilary operators $\hat\phi,\hat\theta,\hat\varsigma$ are given by definitions \eqref{def: theta}-\eqref{def: sigma} and we introduced $\hat\phi^\prime,\hat\varsigma^\prime$ as:
\begin{align}
\hat\phi^\prime_{\tilde p,\tilde p^\prime}&\defeq \hat\phi_{\tilde p,\tilde p^\prime}(\hat n_{\tilde p}+1,\hat n_{\tilde p^\prime}+1) \label{def: phi_prime}\\
\hat\varsigma^\prime_k&\defeq \hat\varsigma_k(\hat  n_k+1). \label{def: sigma_prime}
\end{align}

\onecolumngrid 
\section{Operations on Fourier Series-type Operators}
\label{sec:app-B}
In the flow equations, we frequently encounter expressions of the form $\hat O(\hat n_q\pm1,\hat n_{q^\prime}\pm1)$ which can be evaluated explicitly using addition theorems for sine and cosine. Let $\xi_1,\xi_2\in\{+1,-1\}$. Then:
\begin{align}
&\hat O(\hat n_q+\xi_1,\hat n_{q^\prime}+\xi_2)-\alpha_0^{(O)}\nonumber\\
&= \Sum_{n=1}^N \alpha^{(O)}_n\cos\Big(\underbrace{naq-n\cdot a\hat P}_{\eqdef n\mathcal P }-na(\xi_1 k_{q,x}+\xi_2 k_{q^\prime,x})\Big)\nonumber+ \Sum_{n=1}^N \beta^{(O)}_n\sin\left( n\mathcal P-na(\xi_1 k_{q,x}+\xi_1 k_{q^\prime,x})\right)\nonumber\\
&= \Sum_{n=1}^N \alpha^{(O)}_n\cos\left( n\mathcal P\right)\cos\left(na(\xi_1 k_{q,x}+\xi_2 k_{q^\prime,x})\right)\nonumber+ \Sum_{n=1}^N \alpha^{(O)}_n\sin\left( n\mathcal P\right)\sin\left(na(\xi_1 k_{q,x}+\xi_2 k_{q^\prime,x})\right)\nonumber\\
&+ \Sum_{n=1}^N \beta^{(O)}_n\sin\left( n\mathcal P\right)\cos\left(na(\xi_1 k_{q,x}+\xi_2 k_{q^\prime,x})\right)\nonumber- \Sum_{n=1}^N \beta^{(O)}_n\cos\left( n\mathcal P\right)\sin\left(na(\xi_1 k_{q,x}+\xi_2 k_{q^\prime,x})\right)\nonumber\\
&= \Sum_{n=1}^N \cos\left( n\mathcal P\right)\Big(\alpha^{(O)}_n\cos\left(na(\xi_1 k_{q,x}+\xi_2 k_{q^\prime,x})\right)-\beta^{(O)}_n\sin\left(na(\xi_1 k_{q,x}+\xi_2 k_{q^\prime,x})\right)\nonumber\\
&+ \Sum_{n=1}^N \sin\left( n\mathcal P\right)\Big(\beta^{(O)}_n\cos\left(na(\xi_1 k_{q,x}+\xi_2 k_{q^\prime,x})\right)+\alpha^{(O)}_n\sin\left(na(\xi_1 k_{q,x}+\xi_2 k_{q^\prime,x})\right)\Big) \label{eq: OP1}
\end{align}
Now consider the product of two operators $\hat O,\hat O^\prime$:
\begin{align}
&\hat O\hat O^\prime-\alpha^{(O)}_0\hat O^\prime-\alpha^{(O^\prime)}_0\hat O\nonumber\\
&=\Sum_{n=1}^N \Sum_{{\tilde n}=1}^N\left(\alpha_n^{(H)}\cos\left(n\mathcal P\right)+\beta_n^{(H)}\sin\left(n\mathcal P\right)\right)\left(\alpha_{\tilde n}^{(H^\prime)}\cos\left({\tilde n}\mathcal P\right)+\beta_{\tilde n}^{(H^\prime)}\sin\left({\tilde n}\mathcal P\right)\right)\nonumber\\
&= \Sum_{n=1}^N \Sum_{{\tilde n}=1}^N\Big[\alpha_n^{(H)}\alpha_{\tilde n}^{(H^\prime)}\cos\left(n\mathcal P\right)\cos\left(\tilde n\mathcal P\right)+\alpha_n^{(H)}\beta_{\tilde n}^{(H^\prime)}\cos\left(n\mathcal P\right)\sin\left(\tilde n\mathcal P\right)\nonumber\\
&\quad\quad\quad  +\alpha_{\tilde n}^{(H^\prime)}\beta_n^{(H)}\sin\left(n\mathcal P\right)\cos\left(\tilde n\mathcal P\right) +\beta_{\tilde n}^{(H^\prime)}\beta_n^{(H)}\sin\left(n\mathcal P\right)\sin\left(\tilde n\mathcal P\right)\Big]\nonumber\\
&= \frac12\Sum_{n=1}^N \Sum_{{\tilde n}=1}^N\Big[\alpha_n^{(H)}\alpha_{\tilde n}^{(H^\prime)}\cos\left((n+\tilde n)\mathcal P\right)+\alpha_n^{(H)}\beta_{\tilde n}^{(H^\prime)}\sin\left((n+\tilde n)\mathcal P\right)\nonumber\\
&\quad\quad\quad  +\alpha_{\tilde n}^{(H^\prime)}\beta_n^{(H)} \sin\left((n+\tilde n)\mathcal P\right)  -\beta_{\tilde n}^{(H^\prime)}\beta_n^{(H)}   \cos\left((n+\tilde n)\mathcal P\right) \Big]          \nonumber\\
&+ \frac12\Sum_{n=1}^N \Sum_{{\tilde n}=1}^N\Big[\alpha_n^{(H)}\alpha_{\tilde n}^{(H^\prime)}\cos\left((n-\tilde n)\mathcal P\right)+\alpha_n^{(H)}\beta_{\tilde n}^{(H^\prime)}\sin\left((\tilde n-n)\mathcal P\right)\nonumber\\
&\quad\quad\quad  +\alpha_{\tilde n}^{(H^\prime)}\beta_n^{(H)} \sin\left((n-\tilde n)\mathcal P\right) +\beta_{\tilde n}^{(H^\prime)}\beta_n^{(H)}   \cos\left((n-\tilde n)\mathcal P\right) \Big]        \nonumber\\
&= \frac12\Sum_{m=2}^{2N} \Sum_{{n}=\min\{1,m-N\}}^{\max\{N,m-1\}}\Big[\alpha_n^{(H)}\alpha_{m-n}^{(H^\prime)}\cos\left(m\mathcal P\right)+\alpha_n^{(H)}\beta_{m-n}^{(H^\prime)}\sin\left(m\mathcal P\right)\nonumber\\
&\quad\quad\quad  +\alpha_{m-n}^{(H^\prime)}\beta_n^{(H)}   \sin\left(m\mathcal P\right)    -\beta_{m-n}^{(H^\prime)}\beta_n^{(H)}      \cos\left(m\mathcal P\right)\Big]           \nonumber\\
&+ \frac12\Sum_{m=1-N}^{N-1} \Sum_{n=\mathrm{max}\left(1,1-m\right)}^{\mathrm{min}\left(N,N-m\right)}\Big[\alpha_n^{(H)}\alpha_{m+n}^{(H^\prime)}\cos\left(m\mathcal P\right)+\alpha_n^{(H)}\beta_{m+n}^{(H^\prime)}\sin\left(m\mathcal P\right)\nonumber\\
&\quad\quad\quad  -\alpha_{m+n}^{(H^\prime)}\beta_n^{(H)}  \sin\left(m\mathcal P\right) +\beta_{m+n}^{(H^\prime)}\beta_n^{(H)}      \cos\left(m\mathcal P\right) \Big].      \label{eq: OP2}
\end{align}

\onecolumngrid 
\section{Coherent State Mean Field Theory}
\label{sec:app-C}
For fixed momentum $q$, we consider the Hamiltonian \eqref{eq: ham_q_with_2phon} and apply the variational coherent state ansatz
\begin{align}
|\psi_q^{\mathrm{MF}}\rangle=\prod_k \exp\left(\alpha_k\CR_k-\alpha_k^* \AN_k\right)|0\rangle\eqdef \hat D(\bm{\alpha})|0\rangle.
\end{align}
$\hat D(\bm{\alpha})$ is the unitary displacement operator which acts on bosonic creation and annihilation operators as follows:
\begin{align}
\hat D(\bm{\alpha})^\dagger\AN_k\hat D(\bm{\alpha})&=\AN_k+\alpha_k,\\
\hat D(\bm{\alpha})^\dagger\CR_k\hat D(\bm{\alpha})&=\CR_k+\alpha_k^*.
\end{align}
Using these relations, the variational coherent state ansatz can be straightforwardly applied to the parts of the Hamiltonian which are at most quadratic in creation and annihilation operators while the term $\sim \cos\hat P$ requires more work. 
As \citet{Grusdt_2014_Bloch} have shown, 
\begin{align}
\langle \psi_q^{\mathrm{MF}}|\cos\left(aq-a\hat P\right)|\psi_q^{\mathrm{MF}}\rangle = e^{-C\left[\bm\alpha\right]}\cos\left(aq-S\left[\bm\alpha\right]\right),
\end{align}
where the functionals
\begin{align}
C\left[\bm\alpha\right]&\defeq \int\mathrm dk |\alpha_k|^2\left(1-\cos(ak)\right)\\
S\left[\bm\alpha\right]&\defeq \int\mathrm dk |\alpha_k|^2\sin(ak).
\end{align}
have been introduced.\par
Neglecting the constant terms in $\ham_q$, the MF energy functional becomes
\begin{align} \label{eq: MF_energy_functional}
\hamdens^{\mathrm{MF}}_q\left[\bm\alpha\right]&=\langle \psi_q^{\mathrm{MF}}|\ham_q|\psi_q^{\mathrm{MF}}\rangle =-2Je^{-C\left[\bm\alpha\right]}\cos\left(aq-S\left[\bm\alpha\right]\right) + \int\mathrm dk\left[\omega_k |\alpha_k|^2+U_k(\alpha_k+\alpha_k^*)\right]\\
&+\int\mathrm dk\int\mathrm dk^\prime\left[V_{k,k^\prime}\alpha_k^*\alpha_{k^\prime}+W_{k,k^\prime}\left(\alpha_k\alpha_{k^\prime}+\alpha_k^*\alpha_{k^\prime}^*\right)\right].
\end{align}
Minimizing this energy functional w.r.t $\bm\alpha$ gives, for each k, the following self-consistency equation for the variational parameters:
\begin{align}
\frac{\delta\hamdens^{\mathrm{MF}}_q}{\delta\alpha_k}=0\iff\alpha_k=-\frac{1}{\Omega_k[\bm\alpha]}\left(U_k+\int\mathrm d\tilde k\left[V_{\tilde k,k}\alpha_{\tilde k}+\left(W_{k,\tilde k}+W_{\tilde k,k}\right)\alpha_{\tilde k}^*\right]\right),
\end{align}
where 
\begin{align}
\Omega_k[\bm\alpha]\defeq \omega_k+2Je^{-C\left[\bm\alpha\right]}\left(\cos\left(aq-S\left[\bm\alpha\right]\right)-\cos\left(aq-ak-S\left[\bm\alpha\right]\right)\right).
\end{align}
Instead of solving the self-consistency equation, it is also feasable to directly minimize the energy functional \eqref{eq: MF_energy_functional}. The latter, however, does not produce physical results below a critical (attractive) interaction strength where the energy functional becomes unbounded from below while the self-consistency equation still yields a saddle-point solution. 

\end{document}